\documentclass[11pt]{article}

\include{setup}

\allowdisplaybreaks[1]

\begin{document}

\thispagestyle{empty}
\def\thefootnote{\fnsymbol{footnote}}
\setcounter{footnote}{1}
\null
\draftdate\hfill MPP-2009-73 \\
\strut\hfill PITHA 09/12 \\
\strut\hfill PSI-PR-09-10
\vskip 0cm
\vfill
\begin{center}
  {\Large \boldmath{\bf
      Charged-Higgs-boson production at the LHC:\\[.5em]
      NLO supersymmetric QCD corrections}
    \par} \vskip 2.5em {\large
    {\sc Stefan Dittmaier}\\[1ex]
    {\normalsize \it Physikalisches Institut,
      Albert-Ludwigs-Universit\"at Freiburg, \\
      D--79104 Freiburg, Germany and\\
      Max-Planck-Institut f\"ur Physik (Werner-Heisenberg-Institut),
      F\"ohringer~Ring~6, D--80805~M\"unchen, Germany}\\[2ex]
    {\sc Michael Kr\"amer}\\[1ex]
    {\normalsize \it Institut f\"ur Theoretische Physik, RWTH Aachen University, \\
      D--52056 Aachen, Germany}\\[2ex]
    {\sc Michael Spira}\\[1ex]
    {\normalsize \it Paul Scherrer Institut, CH--5232 Villigen PSI,
      Switzerland}\\[2ex]
    {\sc Manuel Walser}\\[1ex]
    {\normalsize \it Paul Scherrer Institut, CH--5232 Villigen PSI,
      Switzerland and\\ Institute for Theoretical Physics, ETH
      Z\"urich, CH--8093 Z\"urich,
      Switzerland}}\\[2ex]

\par \vskip 1em
\end{center}\par

\vskip .0cm \vfill {\bf Abstract:}\par\noindent The dominant
production process for heavy charged Higgs bosons at the LHC is the
associated production with heavy quarks.  We have calculated the
next-to-leading-order supersymmetric QCD corrections to charged-Higgs
production through the parton processes $\Pq\bar \Pq,\Pg\Pg \to {\rm
  t}\Pb{\rm H}^{\pm}$ and present results for total cross sections and
differential distributions. The QCD corrections reduce the
renormalization and factorization scale dependence and thus stabilize
the theoretical predictions. We present a comparison of the
next-to-leading-order results for the inclusive cross section with a
calculation based on bottom--gluon fusion $\Pg \Pb \to {\rm t}{\rm
  H}^{\pm}$ and discuss the impact of the next-to-leading-order
corrections on charged-Higgs searches at the LHC.
\par
\null
\setcounter{page}{0}
\clearpage
\def\thefootnote{\arabic{footnote}}
\setcounter{footnote}{0}

\section{Introduction}
\label{sec:intro}

The Higgs mechanism~\cite{Higgs:1964ia} is a cornerstone of the
Standard Model (SM) and its supersymmetric extensions. The masses of
the fundamental particles, electroweak gauge bosons, leptons, and
quarks, are generated by interactions with Higgs fields. The search
for Higgs bosons is thus one of the most important tasks for
high-energy physics and is being pursued at the upgraded
proton--antiproton collider Tevatron with a centre-of-mass (CM) energy
of $1.96$~TeV, followed by the proton--proton collider LHC with
$14$~TeV CM energy scheduled to start taking data in 2010.

The minimal supersymmetric extension of the Standard Model (MSSM)
requires two Higgs doublets leading to five physical scalar Higgs
bosons: two (mass-degenerate) charged particles ${\rm H}^\pm$, one
CP-odd neutral particle A, and two CP-even neutral particles h,H. The
discovery of a charged Higgs boson, in particular, would provide
unambiguous evidence for an extended Higgs sector beyond the Standard
Model.  Searches at LEP have set a limit $M_{{\rm H}^\pm} > 79.3$~GeV
on the mass of a charged Higgs boson in a general two-Higgs-doublet
model~\cite{Heister:2002ev}. Within the MSSM, the charged-Higgs mass
is constrained by the pseudoscalar Higgs mass and the W-boson mass
through $M^2_{{\rm H}^\pm} = M^2_{{\rm A}} + M^2_{{\rm W}}$ at tree
level, with only moderate higher-order
corrections~\cite{Gunion:1988pc, Brignole:1991wp, Diaz:1991ki,
  Frank:2006yh}. A mass limit on the MSSM charged Higgs boson can thus
be derived from the limit on the pseudoscalar Higgs boson, $M_{{\rm
    A}} > 93.4$~GeV~\cite{Schael:2006cr}, resulting in $M_{{\rm
    H}^\pm}\gsim 120$~GeV. At the Tevatron, searches for light charged
Higgs bosons in top-quark decays $\Pt \to \Pb {\rm
  H}^\pm$~\cite{Abulencia:2005jd, Abazov:2001md} have placed some
constraints on the MSSM parameter space, but do not provide any
further generic bounds on $M_{\rm H}^{\pm}$.

The LHC will extend the search for charged Higgs bosons to masses up
to $M_{{\rm H}^\pm}\lsim$600~GeV~\cite{:1999fq, Ball:2007zza}, where
the reach depends in detail on the values of the supersymmetric
parameters. In this paper we shall focus on the most promising search
channel for heavy ${\rm H}^\pm$ (with $M_{{\rm H}^\pm} \gsim m_{\rm
  t}$) at the LHC, which is the associated production of charged Higgs
with heavy quarks,
\begin{equation}
  \Pp\Pp \to {\rm t}\Pb{\rm H}^{\pm}+X\,.
\end{equation}
Alternative production mechanisms like quark--antiquark annihilation
$\Pq\bar \Pq'\to \PH^\pm$, $\PH^\pm+\mathrm{jet}$ production,
associated $\PH^\pm\PW^\mp$ production or Higgs pair production have
suppressed rates, and it is not yet clear whether a signal could be
established in any of those channels (see Ref.~\cite{Djouadi:2005gj}
and references therein). Some of the above production processes may,
however, be enhanced in models with non-minimal flavour violation
(see, e.g., \citere{Dittmaier:2007uw}).

Two different formalisms can be employed to calculate the cross
section for associated ${\rm t}\Pb{\rm H}^\pm$ production.  In a
four-flavour scheme (4FS) with no $\Pb$ quarks in the initial state,
the lowest-order QCD production processes are gluon--gluon fusion and
quark--antiquark annihilation, $\Pg\Pg \to {\rm t}\Pb{\rm H}^\pm$ and
$\Pq\bar \Pq \to {\rm t}\Pb{\rm H}^\pm$, respectively.  The inclusive
cross section for $\Pg\Pg \to {\rm t}\Pb{\rm H}^\pm$ develops
potentially large logarithms $\propto \ln(\mu_{\rm F}/m_\Pb)$, which
arise from the splitting of incoming gluons into nearly collinear
$\Pb\bar \Pb$ pairs. The large scale $\mu_{\rm F}$ of ${\cal
  O}(M_{{\rm H}^\pm})$ corresponds to the upper limit of the collinear
region up to which factorization is valid. The $\ln(\mu_{\rm
  F}/m_\Pb)$ terms can be summed to all orders in perturbation theory
by introducing bottom parton densities. This defines the so-called
five-flavour scheme (5FS)~\cite{Barnett:1987jw}. The use of bottom
distribution functions is based on the approximation that the outgoing
$\Pb$ quark is at small transverse momentum and massless, and the
virtual $\Pb$ quark is quasi on-shell. In this scheme, the
leading-order (LO) process for the inclusive ${\rm tbH}^\pm$ cross
section is gluon--bottom fusion, $\Pg \Pb \to {\rm tH}^\pm$.  The
next-to-leading order (NLO) cross section in the 5FS includes ${\cal
  O}(\alpha_{\mathrm{s}})$ corrections to $\Pg \Pb \to {\rm tH}^\pm$
and the tree-level processes $\Pg\Pg \to {\rm tbH}^\pm$ and $\Pq\bar
\Pq \to {\rm tbH}^\pm$.

To all orders in perturbation theory the four- and five-flavour
schemes are identical, but the way of ordering the perturbative
expansion is different, and the results do not match exactly at finite
order. For the inclusive production of neutral Higgs bosons with
bottom quarks, $\Pp\Pp \to \Pb\bar{\Pb}{\rm H}+X$, the four- and
five-flavour scheme calculations numerically agree within their
respective uncertainties, once higher-oder QCD corrections are taken
into account~\cite{Dittmaier:2003ej, Campbell:2004pu, Dawson:2005vi,
  Buttar:2006zd}.  However, no NLO comparison of the 4FS and 5FS
calculations for charged-Higgs production with heavy quarks exists so
far.

There has been considerable progress recently in improving the
cross-section predictions for the associated production of charged
Higgs bosons with heavy quarks by calculating NLO-QCD and SUSY-QCD
corrections in the four and five-flavour schemes~\cite{Zhu:2001nt,
  Gao:2002is, Plehn:2002vy, Berger:2003sm, Kidonakis:2005hc,
  Peng:2006wv}.  The inclusion of higher-order effects is crucial for
an accurate theoretical prediction and, eventually, a determination of
Higgs-boson parameters from the comparison of theory and experiment.
In this paper we present an independent calculation of the NLO
supersymmetric QCD corrections to the process $\Pp\Pp \to {\rm t
  bH}^\pm+X$ in the 4FS. The calculation within the 4FS allows to
describe the dynamics of the final-state bottom quark, which in the
5FS scheme calculation at LO is assumed to be always produced at small
transverse momentum and is thus treated inclusively\footnote{This
  shortcoming of the 5FS, however, is rectified when going to NLO,
  where the process $\Pg\Pg \to {\rm t}{\Pb}{\rm H}^\pm$ contributes
  as part of the real corrections.}.  However, Monte Carlo simulations
show that in about 20\% of $\Pp\Pp\to {\rm tbH}^{\pm}+X$ events at the
LHC the b quark from the production process has a transverse momentum
larger than the b quark from the top-quark decay, and will thus
contaminate the event reconstruction~\cite{kinnunen_2006}.  We
therefore provide state-of-the art NLO predictions not only for the
inclusive cross section but also for various differential
distributions. In contrast to previous analyses our results are based
on the consistent use of a four-flavour parton distribution function.
Furthermore, we present the first comparison of the 4FS and 5FS
calculations at NLO for the inclusive ${\rm tH}^\pm$ cross section.

The paper is organized as follows: In Section~\ref{se:corrections} we
shall describe the calculation of the NLO supersymmetric QCD
corrections.  Numerical results for MSSM Higgs-boson production at the
LHC are presented in Section~\ref{se:pheno}.  We conclude in
Section~\ref{se:conclusion}.  The Appendix provides details on the
scenario of the supersymmetric model under consideration.

\section{Calculation}
\label{se:corrections}

\subsection{LO processes and conventions}

\begin{figure}
\epsfig{file=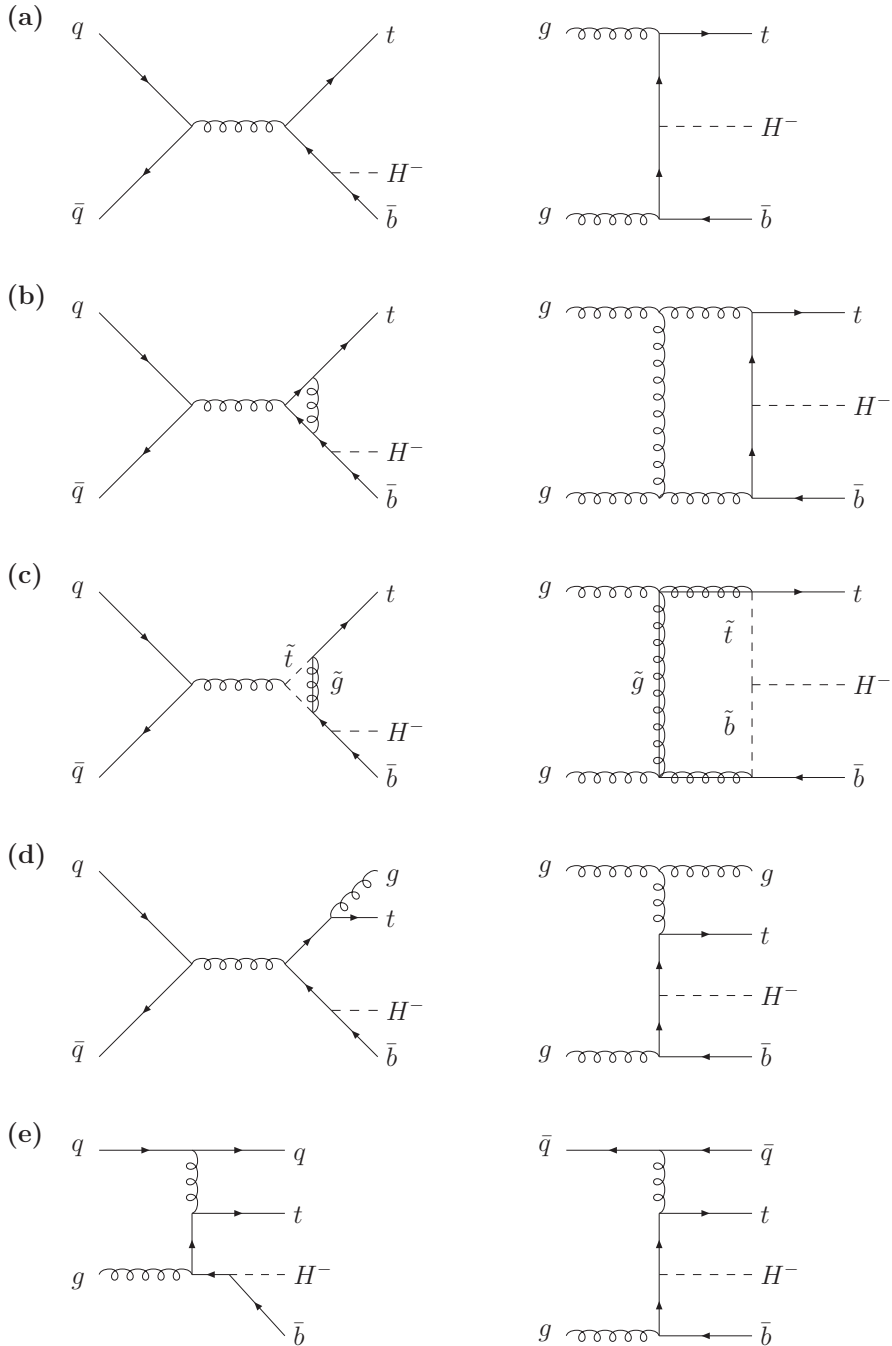,%
        bbllx=0pt,bblly=20pt,bburx=596pt,bbury=842pt,%
        scale=0.7,clip=}
\caption{A generic set of diagrams (a) for the Born level, (b) for
virtual gluon exchange, (c) virtual gluino and squark exchange, (d) gluon
radiation, and (e) gluon--(anti)quark scattering in the subprocesses
$\Pq\bar \Pq, \Pg\Pg\to {\rm t}\bar \Pb{\rm H}^-$, etc.}
\label{fig:diags}
\end{figure}

In the 4FS the production of charged Higgs bosons in association with
top and bottom quarks proceeds at LO through the parton
processes~\cite{DiazCruz:1992gg, Borzumati:1999th, Miller:1999bm}
\begin{equation}
\label{eq:lo_process}
\Pg\Pg \to {\rm t}\bar{\Pb}{\rm H}^- \qquad \mbox{and} \qquad
\Pq\bar{\Pq} \to {\rm t}\bar{\Pb}{\rm H}^-\,,
\end{equation} 
and the charge-conjugate processes with the $\bar{\rm t}{\rm bH}^+$
final state. Throughout this paper we present results for the ${\rm
  t}\bar{\Pb} {\rm H}^-$ channels, unless stated otherwise. Generic
Feynman diagrams that contribute to the LO processes
(\ref{eq:lo_process}) are displayed in Fig.~\ref{fig:diags}(a).

In the MSSM, the Yukawa coupling of the charged Higgs boson ${\rm
  H}^-$ to a top and bottom quark is given by
\begin{equation}
  g_{{\rm t}\bar\Pb{\rm H}^-} = 
  \sqrt{2}\left( \frac{m_{\rm t}}{v} P_R \cot\beta +
    \frac{m_\Pb}{v}  P_L\tan\beta
\right)\,,
\end{equation}
where $v = \sqrt{v_1^2+v_2^2} = (\sqrt2 G_{\rm F})^{-1/2}$ is the
vacuum expectation value of the Higgs field in the Standard Model, and
$G_{\rm F} = 1.16637\times 10^{-5}~$GeV$^{-2}$~\cite{Yao:2006px} is
the Fermi constant.  The ratio of the vacuum expectation values $v_1$
and $v_2$ of the two Higgs doublets is denoted by $\tan\beta =
v_2/v_1$, and $P_{L/R} = (1\mp \gamma_5)/2$ are the chirality
projectors.

\subsection{NLO supersymmetric QCD corrections}
The NLO supersymmetric QCD corrections comprise virtual one-loop
diagrams, Fig.~\ref{fig:diags}(b,c), gluon radiation processes,
Fig.~\ref{fig:diags}(d), and gluon--(anti)quark scattering reactions,
Fig.~\ref{fig:diags}(e). The NLO QCD calculation of the SM processes
$\Pq\bar \Pq,\Pg\Pg \to Q\overline{Q}{\rm H}$, where $Q$ denotes a
generic heavy quark, has been described in some detail in
\citeres{Beenakker:2001rj, Beenakker:2002nc} (see also
Ref.~\cite{Dawson:2002tg}).  Following closely
\citeres{Beenakker:2001rj, Beenakker:2002nc}, we have performed two
independent calculations of the virtual and real corrections, which
are in mutual agreement. A detailed account of one of the two
calculations of the virtual corrections is presented in
Ref.~\cite{diss_walser}. In the following we provide a short summary
of our methods and mention the tools that have been used.

The Feynman diagrams and amplitudes that contribute to the virtual
corrections have been generated with {\sl Feyn\-Arts}~1.0
\cite{Kublbeck:1990xc} and {\sl Feyn\-Arts}~3.2 \cite{Hahn:2000kx}.
The amplitudes have been processed further with two independent
in-house {\sl Mathematica} routines, which automatically create output
in {\sl Fortran} and {\sl C++}, respectively.  The IR (soft and
collinear) singularities have been regularized in $D=4-2\epsilon$
dimensions and have been separated analytically from the finite
remainder as described in \citeres{Beenakker:2002nc,Dittmaier:2003bc}.
This separation also allows for a transparent evaluation of rational
terms that result from $D$-dependent factors multiplying IR
divergences appearing as poles in $\epsilon$; in agreement with the
general arguments given in \citere{Bredenstein:2008zb} we find that
rational terms of IR origin cancel completely.  The pentagon tensor
integrals have been reduced directly to box integrals following
\citere{Denner:2002ii}. This method does not introduce inverse Gram
determinants in the reduction process, thereby avoiding numerical
instabilities in regions where these determinants become small. Box
and lower-point integrals have been reduced to scalar integrals using
the standard Passarino--Veltman technique~\cite{Passarino:1978jh}.
Sufficient numerical stability is already achieved in this way, but
further improvements with the methods of \citere{Denner:2005nn} are in
progress. The scalar integrals, finally, have been calculated either
analytically or using the results of \citeres{'tHooft:1978xw}.  The
IR-finite scalar integrals have furthermore been checked with {\sl
  LoopTools/FF}~\cite{Hahn:1998yk}.

Both evaluations of the real-emission corrections employ (independent
implementations of) the dipole subtraction formalism
\cite{Catani:1996vz} for the extraction of IR singularities and for
their combination with the virtual corrections. Helicity amplitudes
for the real emission processes have been generated and evaluated with
{\sl Madgraph}~\cite{Alwall:2007st} and {\sl
  HELAS}~\cite{Murayama:1992gi}.  The result has been checked by an
independent calculation using standard trace techniques.

\subsection{Parameter renormalization and resummation improvements}

The renormalization of the strong coupling $\alpha_{\mathrm{s}}(\mu)$
and the factorization of initial-state collinear singularities are
performed in the $\overline{\mathrm{MS}}$ scheme. As usual, the top
quark and the SUSY particles are decoupled from the running of
$\alpha_{\mathrm{s}}(\mu)$. In the 4FS calculation presented here,
also the bottom quark is decoupled and the partonic cross section is
calculated using a four-flavour $\alpha_{\rm s}$.  While the top- and
bottom-quark masses are defined on-shell, the $\overline{\mathrm{MS}}$
scheme is adopted for the renormalization of the bottom--Higgs Yukawa
coupling, which is fixed in terms of the corresponding
$\overline{\mathrm{MS}}$ renormalization of the bottom mass.  In order
to sum large logarithmic corrections $\propto \ln(\mu/m_\Pb)$ we
evaluate the Yukawa coupling with the running $\Pb$-quark mass
$\overline{m}_{\Pb}(\mu)$~\cite{Braaten:1980yq}.

The SUSY loop corrections induce a modification of the tree-level
relation between the bottom mass and its Yukawa coupling, which is
enhanced at large
$\tan\beta$~\cite{Hall:1993gn,Hempfling:1993kv,Carena:1994bv,Pierce:1996zz}.
These corrections can be summed to all orders by the replacement
\begin{equation}
\label{eq:replacement}
\frac{\Mb\tan\beta}{v} \;\;\to\;\; \frac{\Mb\tan\beta}{v} 
\,\frac{(1 - \Delta_\Pb/\tan^2\beta)}{(1 + \Delta_\Pb)}
\end{equation}
in the bottom Yukawa coupling~\cite{Carena:1999py,Guasch:2003cv},
where
\begin{equation}
\label{eq:qcd_tanb_enhanced}
\Delta_\Pb = \frac{C_F}{2}\frac{\alpha_{\rm s}}{\pi} \, 
m_{\tilde{\Pg}} \, \mu \, \tan\beta \, 
I(m_{\tilde{\Pb}_1},m_{\tilde{\Pb}_2},m_{\tilde{\Pg}}) \, ,
\end{equation}
with $C_F = 4/3$ and the auxiliary function
\begin{equation}
\label{eq:I}
I(a,b,c) = \frac{1}{(a^2-b^2)(b^2-c^2)(a^2-c^2)} \left(
a^2 b^2 \ln \frac{a^2}{b^2} + 
b^2 c^2 \ln \frac{b^2}{c^2} + 
c^2 a^2 \ln \frac{c^2}{a^2}
\right) \, .
\end{equation}
Here, $\tilde{\Pb}_{1,2}$ are the sbottom mass eigenstates, and
$m_{\tilde{\Pg}}$ is the gluino mass.  The summation formalism can be
extended~\cite{Guasch:2003cv} to include corrections proportional to
the trilinear coupling $A_\Pb$.  However, for the MSSM scenarios under
consideration in this work, these corrections turn out to be small,
and the corresponding summation effects may safely be neglected.

If the LO cross section is expressed in terms of the bottom Yukawa
coupling including the summation of the $\tan\beta$-enhanced
corrections (\ref{eq:replacement}), the corresponding NLO contribution
has to be subtracted from the one-loop SUSY-QCD calculation to avoid
double counting. This subtraction is equivalent to an additional
finite renormalization of the bottom mass according to
\begin{equation}
\frac{{\delta}{m_{\Pb}}}{\Mb} = \Delta_\Pb \, \left(1+\frac{1}{\tan^2\beta}\right)\,.
\end{equation}

As we shall demonstrate in the numerical analysis presented in
Section~\ref{se:pheno}, the SUSY-QCD radiative corrections are indeed
sizeable at large $\tan\beta$.  After summation of the
$\tan\beta$-enhanced terms, however, the remaining one-loop SUSY-QCD
corrections are very small, below the percent level.

\section{Phenomenological analysis}
\label{se:pheno}

In this section we present NLO SUSY-QCD predictions for the production
of heavy charged MSSM Higgs bosons at the LHC. We discuss total cross
sections and differential distributions and compare with the 5FS
calculations at NLO for the inclusive ${\rm tH}^-$ cross section.

\subsection{Input parameters}
Let us first specify the values of the input parameters that enter the
numerical analysis.

\begin{itemize}
  
\item[] {\bf SM and MSSM masses:} The top-quark mass is defined on
  shell and set to $172.6$~GeV~\cite{Group:2008nq}.  For the bottom
  pole mass we adopt the value $m_{\Pb} = 4.6$~GeV, corresponding to a
  \MSbar~mass $\overline{m}_{\Pb}(\overline{m}_{\Pb}) = 4.26$~GeV. The
  bottom pole mass enters the calculation of the matrix elements and
  the phase space, while the Higgs Yukawa coupling is evaluated using
  the running bottom mass.  As for the MSSM parameters, we will focus
  on the benchmark scenario SPS~1b~\cite{Allanach:2002nj} which is
  cha\-rac\-terized by a large value of $\tan\beta = 30$ and a
  correspondingly large associated production cross section ${\rm
    pp}\to {\rm tbH}^{\pm}+X$ at the LHC.  The SPS~1b input parameters
  are specified in Appendix~\ref{app:SPS}.  The MSSM tree-level
  relations are used to determine the squark masses that enter the
  SUSY-QCD corrections. The charged-Higgs mass is calculated from
  $\tan\beta$ and the mass of the pseudoscalar Higgs, $M_{\rm A}$,
  taking into account higher-order corrections up to two loops in the
  effective potential approach~\cite{Carena:1995bx, Haber:1996fp} as
  included in the program {\sl HDECAY}~\cite{Djouadi:1997yw}.  For the
  Higgs mass determination we use a five-flavour $\alpha_{\rm s}$ with
  $\alpha_{\rm s}(\MZ) =0.120$~\cite{Martin:2004ir}.  The top quark,
  the squarks, and the gluino are always decoupled from the running of
  the strong coupling.
  
\item[] {\bf Higgs Yukawa coupling:} The evaluation of the
  bottom--Higgs Yukawa coupling, which involves the running
  $\Pb$-quark mass and the summation of the $\tan\beta$-enhanced
  SUSY-QCD corrections through $\Delta_\Pb$, is also based on a
  five-flavour $\alpha_{\rm s}$ with $\alpha_{\rm s}(\MZ) =0.120$. Our
  default choice for the renormalization scale that enters the
  calculation of the running b-quark mass is the average mass of the
  final-state particles, $\mu = (m_{\rm t} + m_{\Pb} + M_{{\rm
      H}^-})/3$. The scale of $\alpha_{\rm s}$ in the summation factor
  of the Yukawa coupling (cf.\ Eq.~(\ref{eq:qcd_tanb_enhanced})), on
  the other hand, is determined by the masses of the supersymmetric
  particles in the loop and is chosen as $\mu = (m_{\tilde{\Pb}_1} +
  m_{\tilde{\Pb}_2} + m_{\tilde{g}})/3$.  This scale choice for the
  effective short-distance contributions included in the resummed
  bottom Yukawa coupling is justified by the recent NNLO results for
  the $\Delta_{\mathrm{b}}$ corrections \cite{Noth:2008tw}.

\item[] {\bf Hadronic cross section:} Our cross-section calculation is
  defined in the four-flavour scheme, i.e.\ with no $\Pb$ quarks in
  the initial state. Thus, for a consistent evaluation of the hadronic
  cross sections we adopt the MRST four-flavour parton
  distribution functions (pdfs)~\cite{Martin:2006qz}. The partonic
  cross section is calculated using the corresponding four-flavour
  $\alpha_{\rm s}$ with $\Lambda^{(4)} = 0.347$~GeV at NLO, except for
  the Higgs Yukawa coupling which is evaluated with a five-flavour
  $\alpha_{\rm s}$ as explained above. Our default choice for the
  renormalization and factorization scales that enter the partonic
  cross section and the pdfs is $\mu = (m_{\rm t} + m_{\Pb} + M_{{\rm
      H}^-})/3$. Note that the LO cross-section predictions have been
  obtained by using the corresponding LO four-flavour pdf set, a LO
  $\alpha_{\rm s}$ with $\Lambda^{(4)} = 0.220$~GeV for the partonic
  cross section and a LO running $\Pb$-quark mass using a LO
  five-flavour $\alpha_{\rm s}$ with $\alpha_{\rm s}(\MZ)
  =0.130$~\cite{Martin:2006qz}.

\end{itemize}

\subsection{Total cross section and scale dependence}

\begin{figure}
\epsfig{file=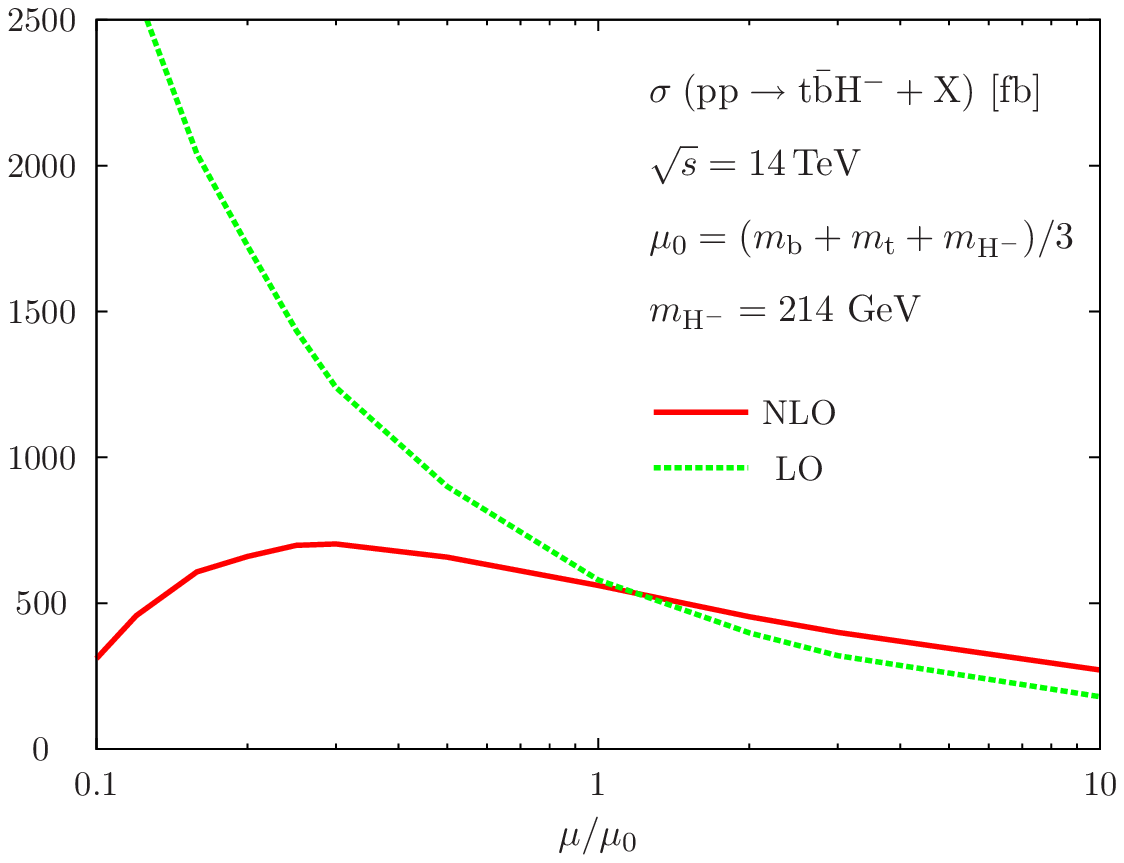,%
        bbllx=170pt,bblly=540pt,bburx=500pt,bbury=820pt,%
        scale=0.69,clip=}
\epsfig{file=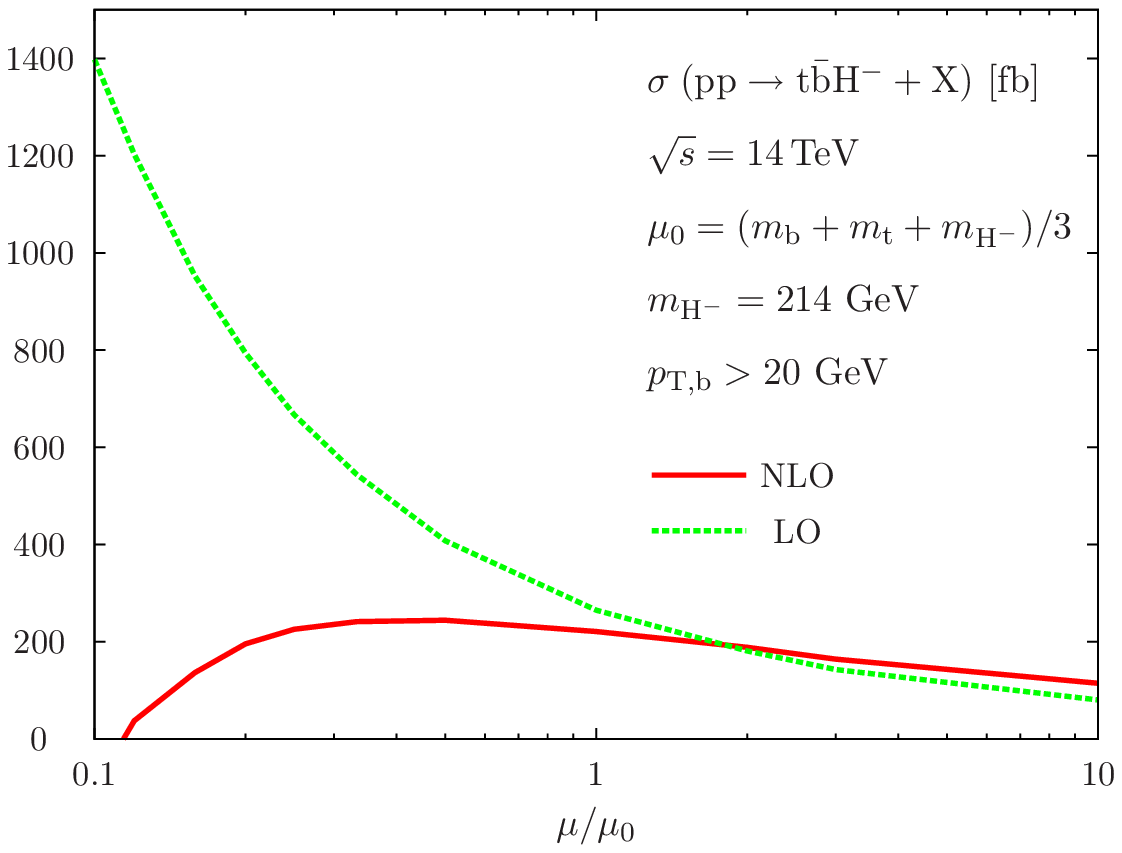,%
        bbllx=170pt,bblly=540pt,bburx=500pt,bbury=820pt,%
        scale=0.69,clip=}
\caption{Variation of the LO and NLO cross sections with the
  renormalization and factorization scales for $\Pp\Pp \to {\rm
    t}\bar{\Pb}{\rm H}^-+X$ at the LHC, without (l.h.s.) and with (r.h.s.) 
    a cut of $p_{\rm T,b} > 20$~GeV on the b-quark transverse momentum.}
\label{fig:scale}
\end{figure}

We first discuss the scale dependence of the total $\Pp\Pp \to {\rm
  t}\bar{\Pb}{\rm H}^-+X$ cross section at the LHC. Note that in NLO
QCD the cross section for the charge-conjugate process $\Pp\Pp \to
\bar{{\rm t}}\Pb{\rm H}^++X$ at the LHC is identical to $\Pp\Pp \to
{\rm t}\bar{\Pb}{\rm H}^-+X$ and can be included by multiplying the
results presented below by a factor of two. The renormalization and
factorization scales that enter the hadronic cross section and the
running b-quark mass are identified and varied around the central
value $\mu_0 = (m_{\rm t} + m_{\Pb} + M_{{\rm H}^-})/3$, but the scale
of $\alpha_{\rm s}$ in the summation factor of the Yukawa coupling
(cf.\ Eq.~(\ref{eq:qcd_tanb_enhanced})), on the other hand, is kept
fixed.  Figure~\ref{fig:scale} (l.h.s.)  shows the scale dependence of
the LO and complete NLO SUSY-QCD cross sections at the SPS~1b
benchmark point and $M_{\rm A} = 200$~GeV, corresponding to $M_{\rm
  H^\pm} = 214.28$~GeV.
As anticipated, the scale dependence of the theoretical prediction is
significantly reduced at NLO, with a remaining uncertainty of
approximately $\pm 25$\% when $\mu$ is varied between $\mu_0/3$ and
$3\mu_0$, compared to approximately $\pm 100$\% at LO.  At the central
scale, the K-factor, ${\rm K} = \sigma_{\rm NLO}/\sigma_{\rm LO}$ is
close to one. Note, however, that the K-factor strongly depends on the
definition of the LO cross section.  As described above, our LO cross
section prediction includes the summation of a certain class of QCD
corrections through a running Yukawa coupling, and has been evaluated
using LO pdfs and $\alpha_{\rm s}$. The reduction of the spurious
scale dependence at NLO is particularly striking for the exclusive
cross section where the $\Pb$ quark is required to be produced with
$p_{\rm T,b} > 20$~GeV, see the r.h.s.\ of Figure~\ref{fig:scale}.
 The QCD corrections for the exclusive cross section are moderate
and negative at the central scale, with a corresponding K-factor of
${\rm K} \approx 0.85$.

\begin{figure}
\epsfig{file=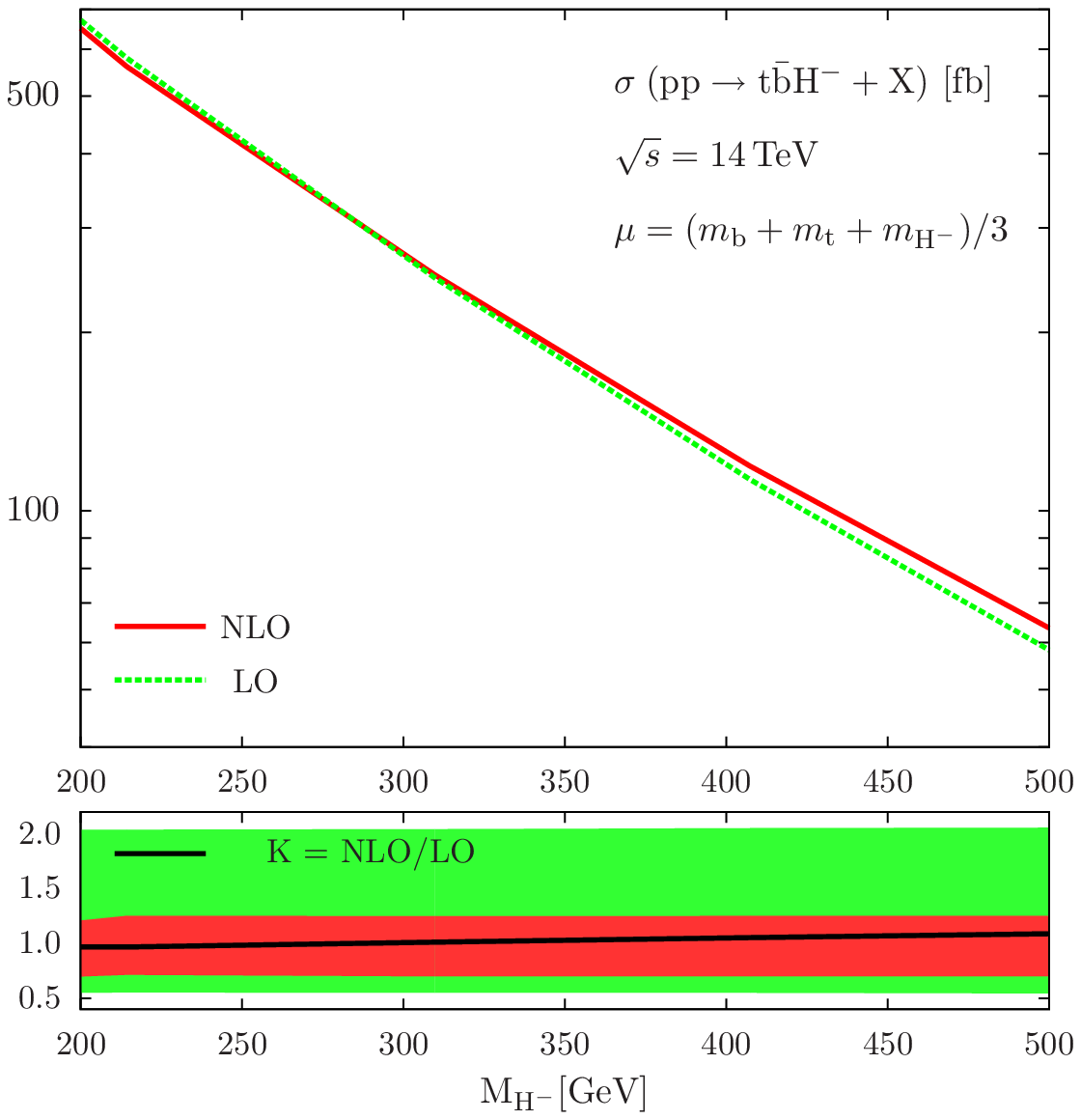,%
        bbllx=170pt,bblly=450pt,bburx=500pt,bbury=820pt,%
        scale=0.69,clip=}
\epsfig{file=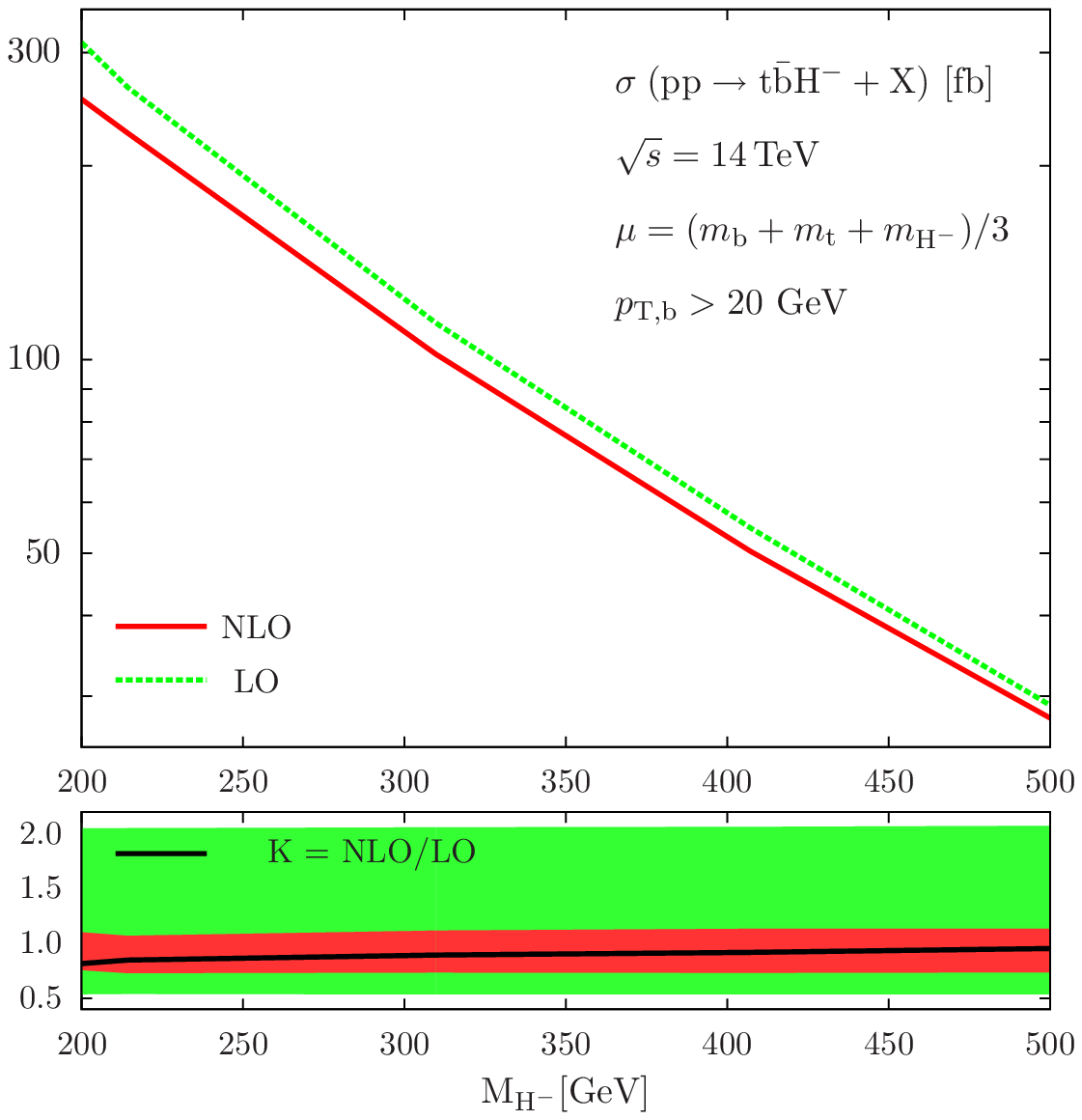,%
        bbllx=170pt,bblly=450pt,bburx=500pt,bbury=820pt,%
        scale=0.69,clip=}
\caption{Total LO and NLO cross sections for $\Pp\Pp \to {\rm
    t}\bar{\Pb}{\rm H}^-+X$ at the LHC as a function of the
  Higgs-boson mass, without (l.h.s.) and with (r.h.s.) 
    a cut of $p_{\rm T,b} > 20$~GeV on the b-quark transverse momentum.
  The lower plots show the K-factor, ${\rm K} =
  \sigma_{\rm NLO}/\sigma_{\rm LO}$, and the scale dependence of the LO
  and NLO cross section predictions for $\mu_0/3 < \mu < 3\mu_0$.}
\label{fig:cxn}
\end{figure}

The total LO and NLO SUSY-QCD cross sections for $\Pp\Pp \to {\rm
  t}\bar{\Pb}{\rm H}^-+X$ at the LHC are shown on the l.h.s.\ of 
Figure~\ref{fig:cxn} as a function of the Higgs-boson mass. 
Note that ${\rm t}\bar{\Pb}{\rm
  H}^-$ production at the LHC is dominated by gluon-induced processes
which provide more than 95\% of the cross section. The K-factor
is displayed in the lower
part of the Figure, together with the scale dependence of the LO and
NLO predictions. We observe that the K-factor is moderate over the
whole range of Higgs-boson masses, with a small increase from ${\rm K}
= 0.97$ at $M_{\rm H^\pm} = 200$~GeV to ${\rm K} = 1.1$ at $M_{\rm
  H^\pm} = 500$~GeV, and that the scale dependence is reduced at NLO
also for large Higgs-boson masses.  Representative values for the
total cross section are listed in Table~\ref{tab:sigma_inclusive}. 
\begin{table}
\begin{center}
\begin{tabular}{|c|c|c|c|c|c|}
\hline
      &            &   & 
\multicolumn{2}{c|}{\rule[-3mm]{0mm}{8mm}
                          $\sigma({\rm pp} \to \bar{\rm t}{\rm
                            bH}^-+X)$ [fb]} & \\
\cline{4-5}
\raisebox{1.9ex}[-1.9ex]{$M_{\rm A}$~[GeV]} &
\raisebox{1.9ex}[-1.9ex]{$M_{{\rm H}^{\pm}}$~[GeV]} & 
\raisebox{1.9ex}[-1.9ex]{$\overline{m}^{\rm NLO}_{\Pb}(\mu)$~[GeV]} &
LO & NLO & 
\raisebox{1.9ex}[-1.9ex]{K $=
\sigma_{\rm NLO}/\sigma_{\rm LO}$}\\
\hline
200 & 214.28 & 2.80 & 583  & 562(2)  & 0.96 \\\hline
300 & 309.70 & 2.76 & 248  & 251(1)  & 1.01 \\\hline
400 & 407.33 & 2.72 & 114  & 119(1)  & 1.04 \\\hline
500 & 505.88 & 2.68 & 56.5 & 61.0(2) & 1.09 \\\hline
\end{tabular}
\end{center}
\mycaption{\label{tab:sigma_inclusive} 
  Total cross sections and K-factors
  for $\Pp\Pp \to {\rm
    t}\bar{\Pb}{\rm H}^-+X$ at the LHC. The renormalization and factorization
  scales are set to $\mu = (m_{\rm t} + m_{\Pb} +
  M_{{\rm H}^-})/3$. The error from the Monte Carlo integration on the
  last digit is given in parenthesis if significant. The MRST four-flavour 
  pdfs~\cite{Martin:2006qz} are adopted. In the 
  third column we show the running b-quark mass
  evaluated at the default renormalization scale.}
\end{table}
To facilitate the comparison with other calculations we also show in
Table~\ref{tab:sigma_inclusive} the running b-quark mass, which enters
the Higgs Yukawa coupling and thus strongly affects the overall
normalization of the cross section.  Requiring the bottom quark to be
produced with $p_{\rm T,b} > 20$~GeV reduces the inclusive cross
section by approximately 60\%, see r.h.s.\ of Figure~\ref{fig:cxn}.  A
systematic comparison of our calculation with the results for the
exclusive cross section with $p_{\rm T,b} > 20$~GeV and $|\eta_{\rm
  b}| < 2.5$ presented in Ref.~\cite{Peng:2006wv} is in progress.

If we adopt -- inconsistently -- the five-flavour pdf
MRST2004~\cite{Martin:2004ir}, on which the four-flavour set is based,
the cross section decreases by approximately 10\%: gluon splitting
into bottom-quark pairs is included in the evolution of the
five-flavour pdf and depletes the gluon flux compared to the
four-flavour pdf. Note that the recent fixed-flavour parton densities
of Ref.~\cite{Gluck:2008gs} are based on three active flavours in the
proton and five active flavours in the evolution of $\alpha_{\rm s}$;
we can thus not use the pdf set of Ref.~\cite{Gluck:2008gs} without
modification of our calculation.

In Table~\ref{tab:sigma_inclusiveII} we show the individual
contributions to the NLO cross section due to the Standard Model QCD
corrections and the genuine SUSY-QCD effects, split further into the
impact of the $\tan\beta$-enhanced corrections included in the
summation factor $\Delta_\Pb$ and the remainder of the genuine SUSY
contributions.
\begin{table}
\begin{center}
\begin{tabular}{|c|c|c|c|c|c|}
\hline
                  & 
\multicolumn{4}{c|}{\rule[-3mm]{0mm}{8mm}
  $\sigma_{\rm NLO} = \sigma_0\!
  \times\! (1\!+\!\delta_{\rm SUSY}^{\tan\beta-{\rm resum.}}) 
 \! \times\!(1\!+\!\delta_{\rm QCD}\! +\! \delta_{\rm SUSY}^{\rm remainder})
                          $} & \\
\cline{2-5}
\raisebox{1.9ex}[-1.9ex]{$M_{{\rm H}^{\pm}}$~[GeV]} & $\sigma_0$ [fb] & 
$\delta_{\rm QCD}$ & $\delta_{\rm SUSY}^{\tan\beta-{\rm resum.}}$
& $\delta_{\rm SUSY}^{\rm remainder}$ &
\raisebox{1.9ex}[-1.9ex]{$\sigma_{\rm NLO}^{\rm fixed-order}$~[fb]} \\
\hline
$214.28$ & $512$  & $0.55$ & $-0.30$ & $-0.0008$ & 562(2)\\\hline
$309.70$ & $224$  & $0.61$ & $-0.30$ & $-0.0012$ & 258(1) \\\hline
$407.33$ & $106$  & $0.61$ & $-0.30$ & $-0.0009$ & 125(1) \\\hline
$505.88$ & $53.3$ & $0.62$ & $-0.30$ & $-0.0002$ & 64.1(2) \\\hline
\end{tabular}
\end{center}
\mycaption{\label{tab:sigma_inclusiveII} 
  LO total cross section $\sigma_0$ and NLO corrections
  $\delta$ relative to $\sigma_0$ for $\Pp\Pp \to {\rm
    t}\bar{\Pb}{\rm H}^-+X$ at the LHC. The error from the Monte Carlo
  integration on the
  last digit is given in parenthesis if significant. The MRST
  pdfs~\cite{Martin:2006qz} are adopted and the renormalization 
  and factorization
  scales have been set to $\mu = (m_{\rm t} + m_{\Pb} +
  M_{{\rm H}^-})/3$. ``QCD'' denotes the NLO QCD corrections only, 
  ``SUSY/$\tan\beta$--resum.'' the
  $\tan\beta$-enhanced SUSY corrections, ``SUSY/remainder''
  the remaining one-loop SUSY corrections and ``NLO/fixed-order''
  the complete NLO  calculation without summation of the
  $\tan\beta$-enhanced 
  terms.}
\end{table}
The cross section labeled $\sigma_0$ denotes the LO parton cross
section evaluated with NLO running $\Pb$-quark mass, pdfs and
$\alpha_{\rm s}$.  The NLO Standard Model QCD corrections,
$\delta_{\rm QCD}$, increase the prediction by approximately $60\%$,
nearly independent of the value of the Higgs-boson mass. This increase
is partially compensated by the $\tan\beta$-enhanced SUSY corrections,
$\delta_{\rm SUSY}^{\tan\beta-{\rm resum.}}$, which amount to
approximately $-30\%$. The impact of the remaining one-loop SUSY-QCD
corrections, $\delta_{\rm SUSY}^{\rm remainder}$, is marginal, below
the percent level. We also show the result of a fixed-order SUSY-QCD
calculation, $\sigma_{\rm NLO}^{\rm fixed-order}$, which does not
include the $\tan\beta$-enhanced corrections beyond NLO. We find that
the effect of the $\tan\beta$-summation beyond NLO, included in our
best cross-section prediction $\sigma_{\rm NLO}$, is moderate, at the
level of $10\%$.

Supersymmetric electroweak ${\cal O(\alpha)}$ corrections have been
studied in Ref.~\cite{Dittmaier:2006cz} for the related process of
neutral MSSM Higgs production in bottom-quark fusion. It has been
shown that the leading corrections can be taken into account by an
appropriate definition of the couplings and the running $\Pb$-mass in
an improved Born approximation. The remaining non-universal
corrections have been found to be small, typically of the order of a
few percent.  It would be interesting to see whether similar
conclusions also hold for the process of charged-Higgs production
studied here.

\subsection{Differential distributions}
\begin{figure}
\epsfig{file=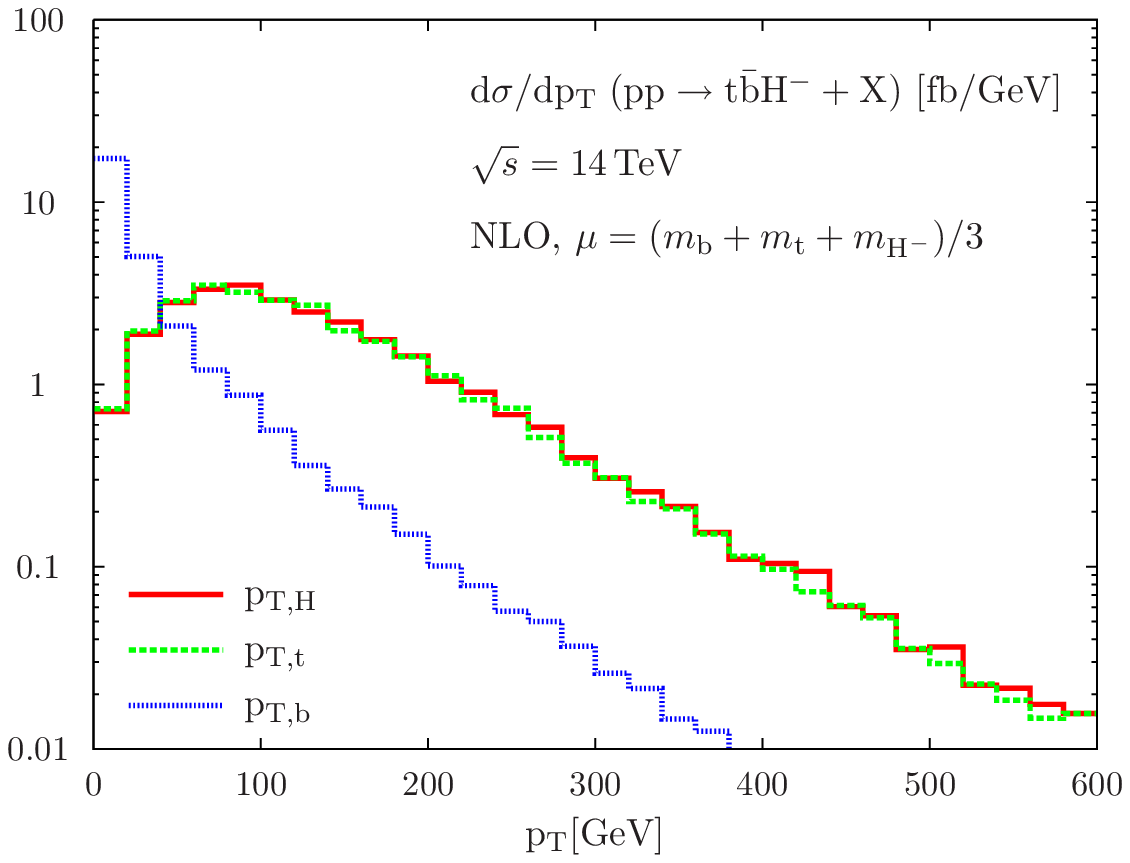,%
        bbllx=170pt,bblly=540pt,bburx=500pt,bbury=820pt,%
        scale=0.69,clip=}
\epsfig{file=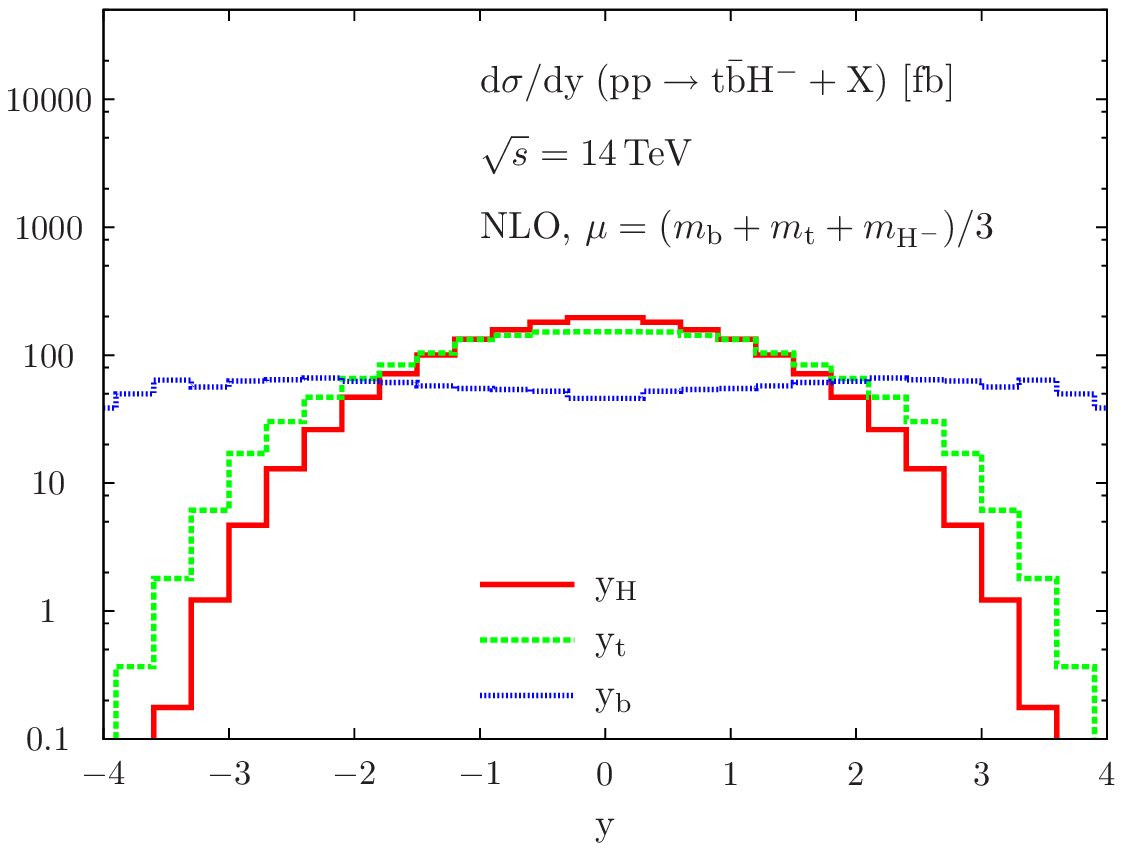,%
        bbllx=170pt,bblly=540pt,bburx=500pt,bbury=820pt,%
        scale=0.69,clip=}
\caption{NLO transverse-momentum and rapidity distributions
  of the Higgs boson, the
  top quark, and the bottom quark for $\Pp\Pp \to {\rm t}\bar{\Pb}{\rm
    H}^-+X$ at the LHC.}
\label{fig:pt_y_comparison}
\end{figure}

Let us now turn to the transverse-momentum and rapidity distributions
of the final-state particles shown in Figure~\ref{fig:pt_y_comparison}.
The distributions have been evaluated for the
default scale choice $\mu = (m_{\rm t} + m_{\Pb} + M_{{\rm H}^-})/3$.
The $p_{\rm T}$-distributions of the top quark and the Higgs boson are
rather similar, with a maximum at $p_{\rm T} \approx 100$~GeV. The
transverse-momentum distribution of the bottom quark is much softer
with $\sigma_{\rm NLO}(p_{\rm T,b}<25~{\rm GeV})/\sigma_{\rm NLO}
\approx 0.7$. The heavy particles, i.e.\  the top quark and the
Higgs boson, are preferentially produced at central rapidities with
$|y|\lsim 2.5$, while the rapidity distribution of the bottom quark is
rather flat in the region $|y| \lsim 4$.

\begin{figure}
\epsfig{file=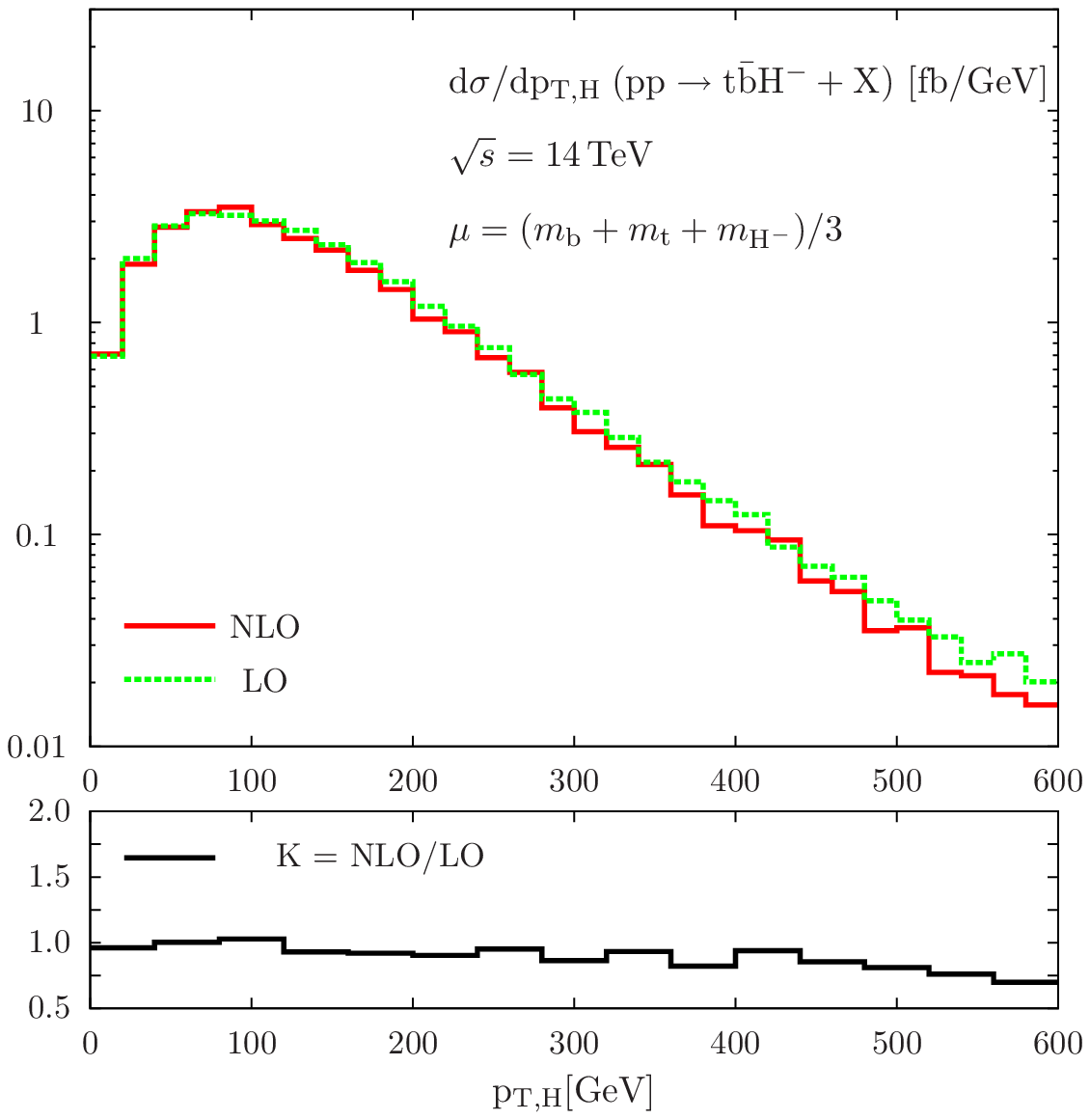,%
        bbllx=170pt,bblly=450pt,bburx=500pt,bbury=820pt,%
        scale=0.69,clip=}
\epsfig{file=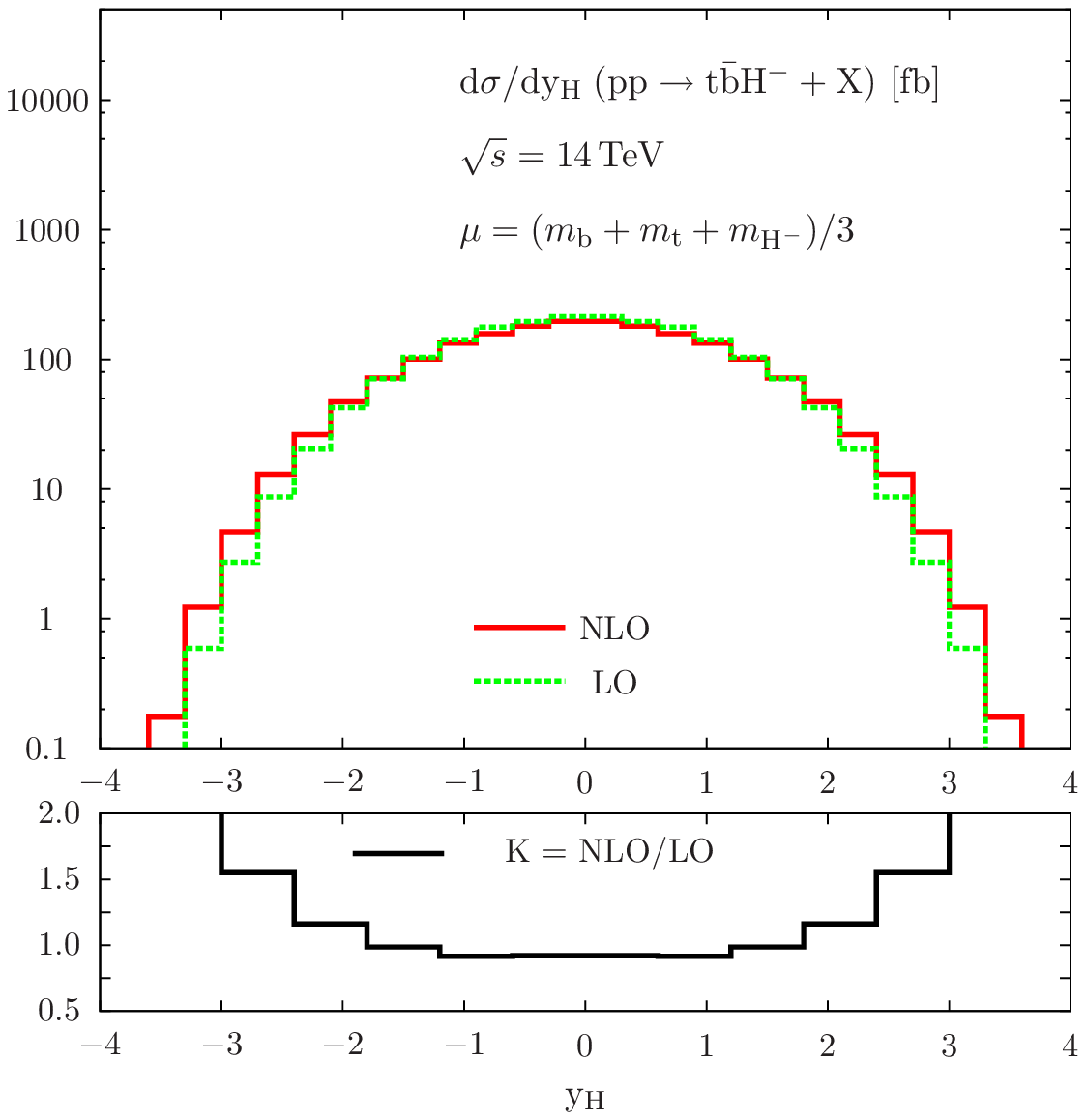,%
        bbllx=170pt,bblly=450pt,bburx=500pt,bbury=820pt,%
        scale=0.69,clip=}
\caption{LO and NLO transverse-momentum and rapidity distributions
  of the Higgs
  boson for $\Pp\Pp \to {\rm t}\bar{\Pb}{\rm H}^-+X$ at the LHC. The
  lower plot shows the K-factor, ${\rm K} = \sigma_{\rm
    NLO}/\sigma_{\rm LO}$.}
\label{fig:ptH_yH}
\end{figure}

\begin{figure}
\epsfig{file=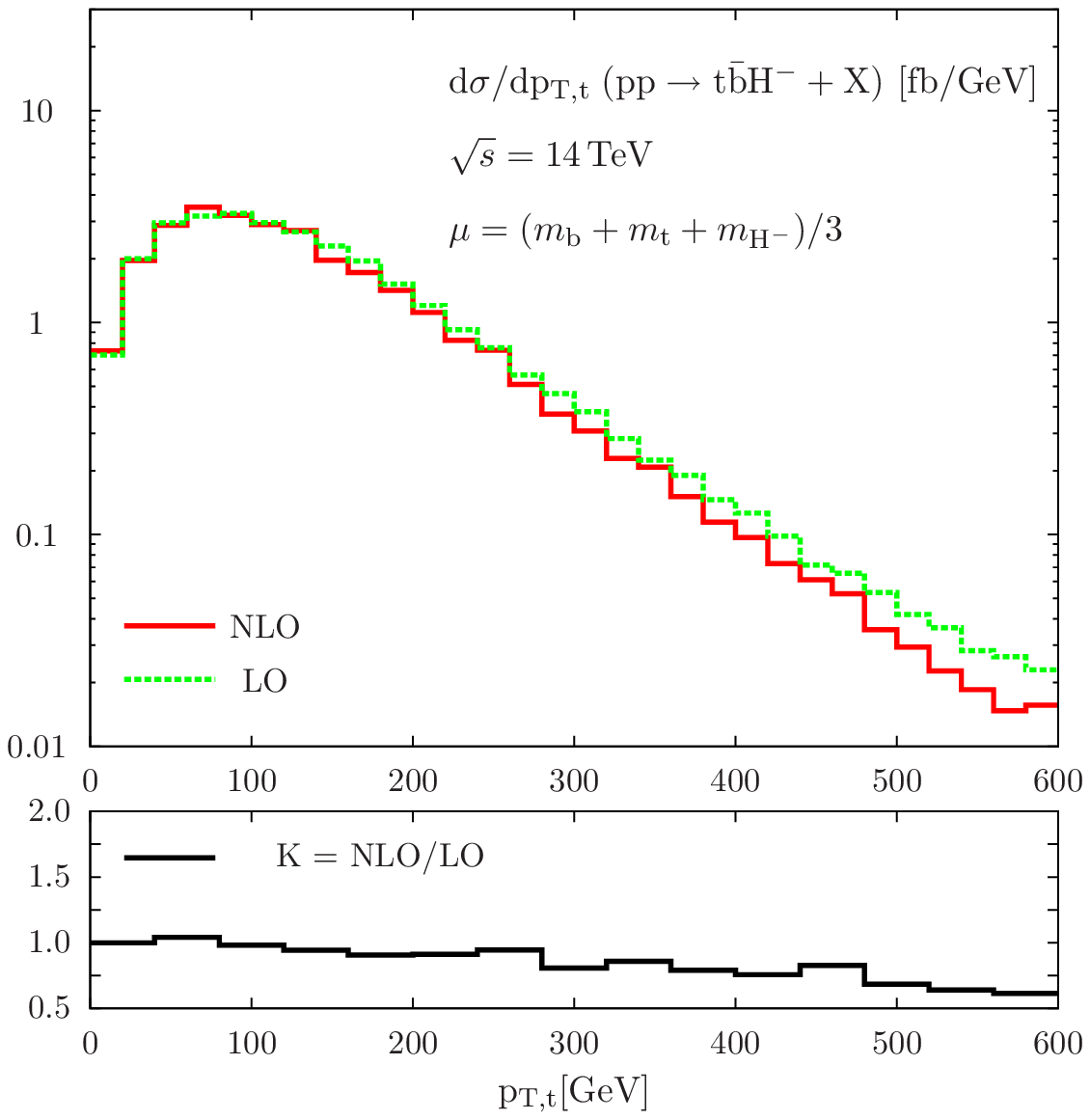,%
        bbllx=170pt,bblly=450pt,bburx=500pt,bbury=820pt,%
        scale=0.69,clip=}
\epsfig{file=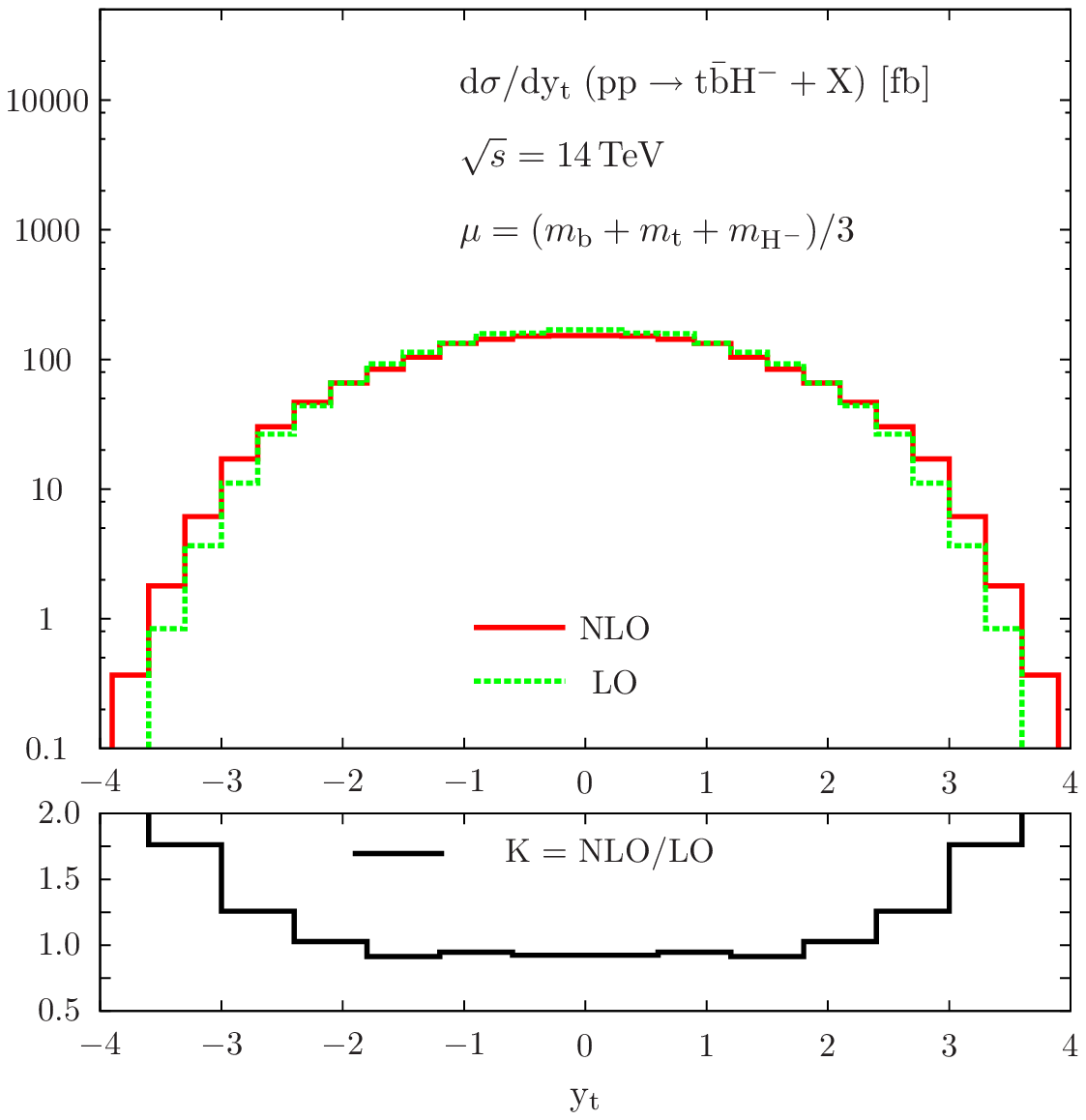,%
        bbllx=170pt,bblly=450pt,bburx=500pt,bbury=820pt,%
        scale=0.69,clip=}
\caption{LO and NLO transverse-momentum and rapidity distributions
  of the top quark
  for $\Pp\Pp \to {\rm t}\bar{\Pb}{\rm H}^-+X$ at the LHC. The lower
  plot shows the K-factor, ${\rm K} = \sigma_{\rm NLO}/\sigma_{\rm
    LO}$.}
\label{fig:ptt_yt}
\end{figure}

\begin{figure}
\epsfig{file=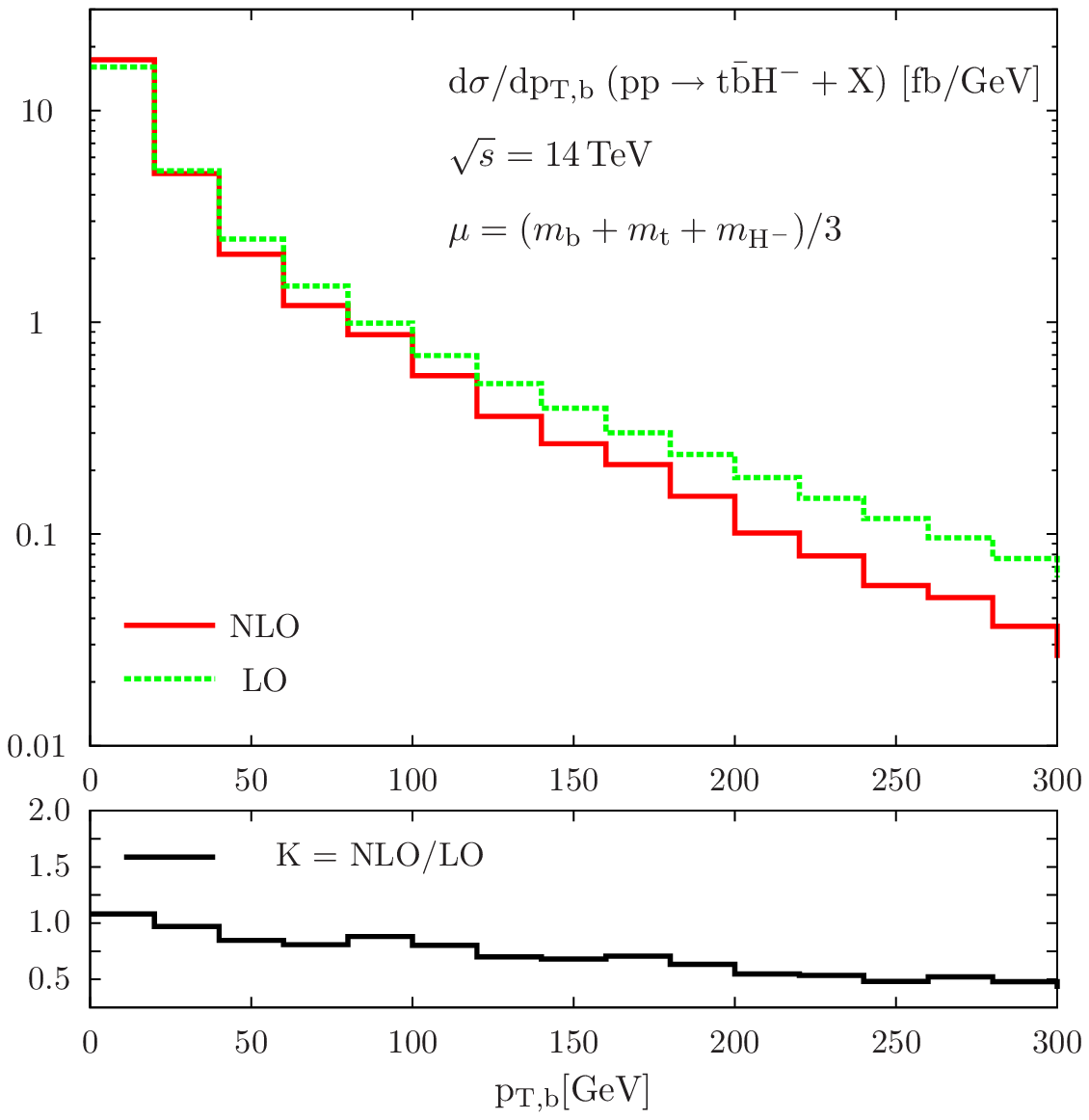,%
        bbllx=170pt,bblly=450pt,bburx=500pt,bbury=820pt,%
        scale=0.69,clip=}
\epsfig{file=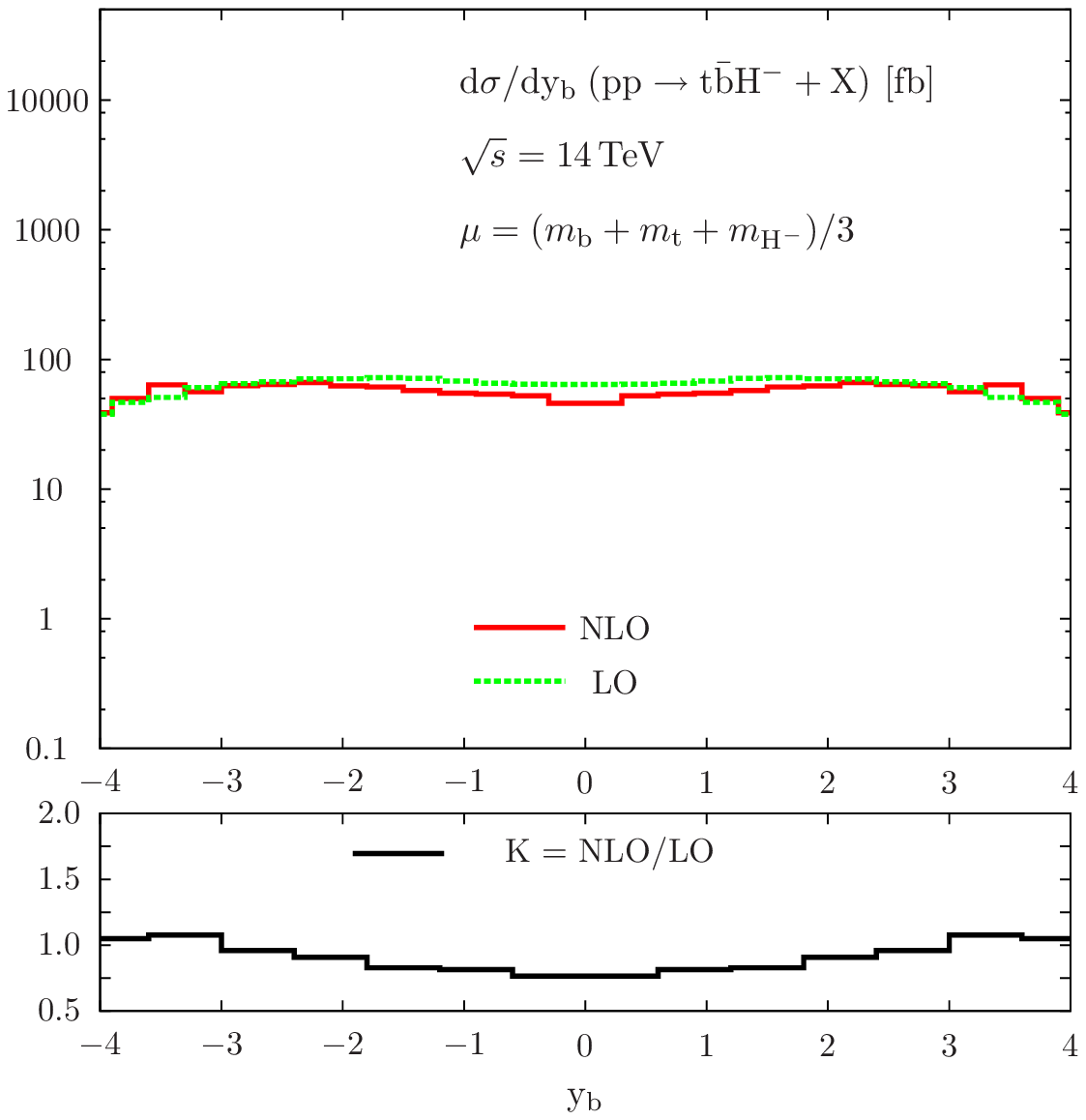,%
        bbllx=170pt,bblly=450pt,bburx=500pt,bbury=820pt,%
        scale=0.69,clip=}
\caption{LO and NLO transverse-momentum and rapidity distributions
  of the bottom
  quark for $\Pp\Pp \to {\rm t}\bar{\Pb}{\rm H}^-+X$ at the LHC. The
  lower plot shows the K-factor, ${\rm K} = \sigma_{\rm
    NLO}/\sigma_{\rm LO}$.}
\label{fig:ptb_yb}
\end{figure}

The impact of the higher-order corrections on the shape of the
transverse-momentum and rapidity distributions is shown in
Figures~\ref{fig:ptH_yH}, \ref{fig:ptt_yt}, and \ref{fig:ptb_yb}, 
respectively.  
The lower part of the Figures shows the K-factor. We find that the
shape of the top and Higgs transverse-momentum distribution is not
strongly affected by the higher-order corrections in the range of
$p_{\rm T}$ relevant for the experimental analysis. On the other hand,
the bottom quark $p_{\rm T}$-distribution, which extends to $p_{\rm
  T,b} \gg m_{\Pb}$, is softened at NLO, with the K-factor decreasing
from ${\rm K} = 1.1$ at $p_{\rm T,b} \approx 20$~GeV to ${\rm K} =
0.5$ at $p_{\rm T,b} \approx 300$~GeV.  The large impact on the
$p_{\rm T,b}$ distribution is due to collinear gluon radiation off
bottom quarks that is enhanced by a factor
$\alpha_{\mathrm{s}}\ln(\Mb/p_{\rm T,b})$. The enhancement should be
significantly reduced if the bottom quarks are reconstructed from
jets, since the application of a jet algorithm treats the
bottom--gluon system inclusively in the collinear cone, so that the
logarithmic enhancement cancels out. The NLO corrections do not
significantly change the shape of the rapidity distributions at
central rapidities $|y|\lsim 2$, but generically predict more
particles at large rapidities $|y|\gsim 2$.

We have also evaluated the differential distributions with the
renormalization and factorization scales set to the average transverse
mass, $\mu = (m_{\rm T,b}+m_{\rm T,t}+m_{\rm T,H})/3$, where $m_{\rm
  T,b} = \sqrt{m_{\rm b}^2 + p_{\rm T,b}^2}$, etc. We find that the
shapes of the NLO distributions are not significantly affected by such
a change. The LO transverse-momentum distributions, however, do
provide a better description of the NLO shapes when evaluated with
$\mu = (m_{\rm T,b}+m_{\rm T,t}+m_{\rm T,H})/3$.

\subsection{Comparison with the 5FS calculation}
\begin{figure}
\epsfig{file=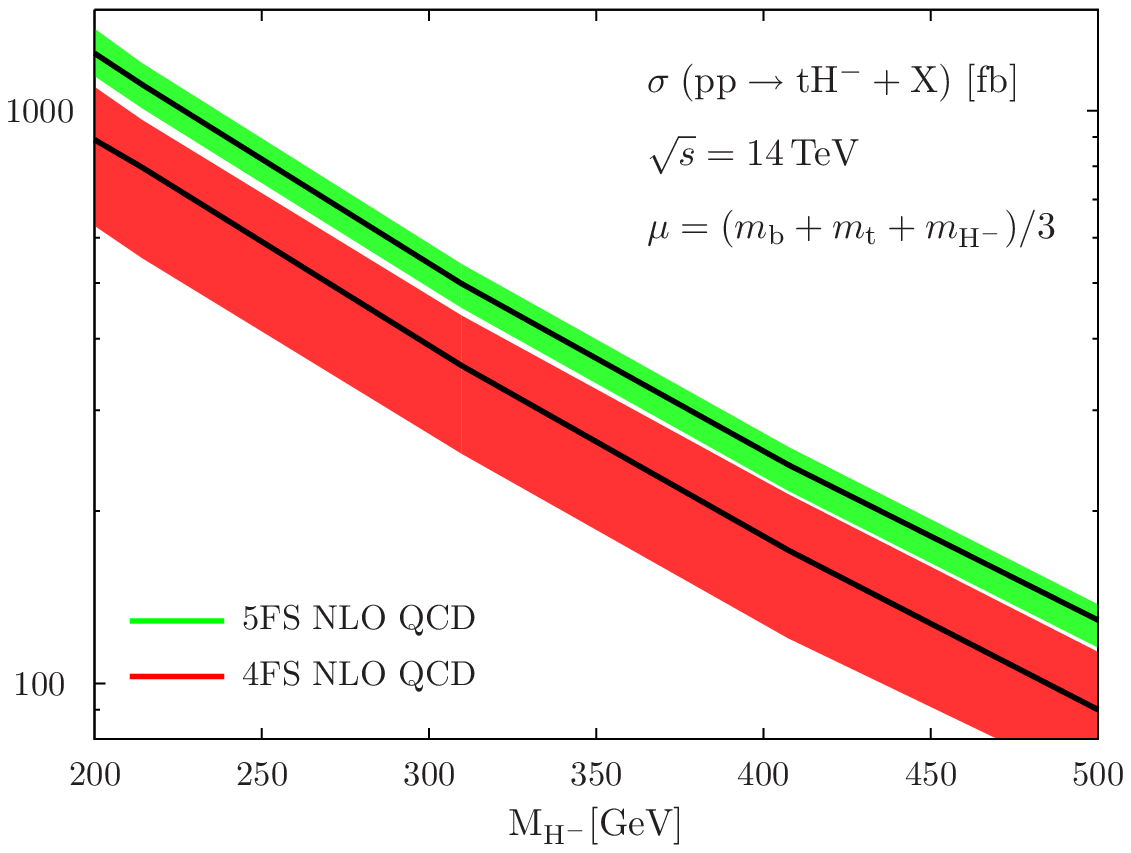,%
        bbllx=170pt,bblly=540pt,bburx=500pt,bbury=800pt,%
        scale=1.0,clip=}
\caption{Total NLO cross section for $\Pp\Pp \to {\rm
    t}{\rm H}^-+X$ at the LHC as
  a function of the Higgs-boson mass in the 4FS and the 5FS. Shown
  is the central prediction and the scale dependence for
  $\mu_0/3 < \mu < 3\mu_0$.}
\label{fig:5fns}
\end{figure}

As discussed in Section~\ref{sec:intro}, in the 5FS the LO process for
the inclusive ${\rm tH}^\pm$ cross section is gluon--bottom fusion,
$\Pg \Pb \to {\rm tH}^\pm$.  The NLO cross section includes ${\cal
  O}(\alpha_{\mathrm{s}})$ corrections to $\Pg \Pb \to {\rm tH}^\pm$
and the tree-level processes $\Pg\Pg \to {\rm tbH}^\pm$ and $\Pq\bar
\Pq \to {\rm tbH}^\pm$, and has been calculated in
Refs.~\cite{Plehn:2002vy, Berger:2003sm}. In Figure~\ref{fig:5fns} we
present a comparison of the 4FS and 5FS calculations at NLO QCD for
the inclusive $\Pp \Pp \to {\rm tH}^{-}+X$ cross section at the LHC.
The 5FS calculation is taken from Ref.~\cite{Plehn:2002vy} and is
evaluated with the five-flavour pdf MRST2004~\cite{Martin:2004ir} and
the set of input parameters described above. In particular, the
renormalization and factorization scales have been set to $\mu_0 =
(m_{\rm t} + m_{\Pb} + M_{{\rm H}^-})/3$, as in the 4FS calculation.
The error band indicates the theoretical uncertainty when the
renormalization and factorization scales are varied between $\mu_0/3$
and $3\mu_0$. Thus, the error band also includes the scale choice
$\mu_{\rm F} = (m_{\rm t} + M_{{\rm H}^-})/5$ for the 5FS calculation
advocated in Refs.~\cite{Plehn:2002vy, Berger:2003sm}.  Note that the
cross sections shown in Figure~\ref{fig:5fns} do not include the NLO
SUSY effects, which can be incorporated within good precision by
simply adjusting the bottom Yukawa coupling according to
Eq.~(\ref{eq:replacement}). Even taking the scale uncertainty into
account, the 4FS and 5FS cross sections at NLO are barely consistent;
the central predictions in the 5FS are larger than those of the 4FS by
approximately 40\%, rather independent of the Higgs-boson mass.  

\subsection{Discovery reach}
\begin{figure}
\epsfig{file=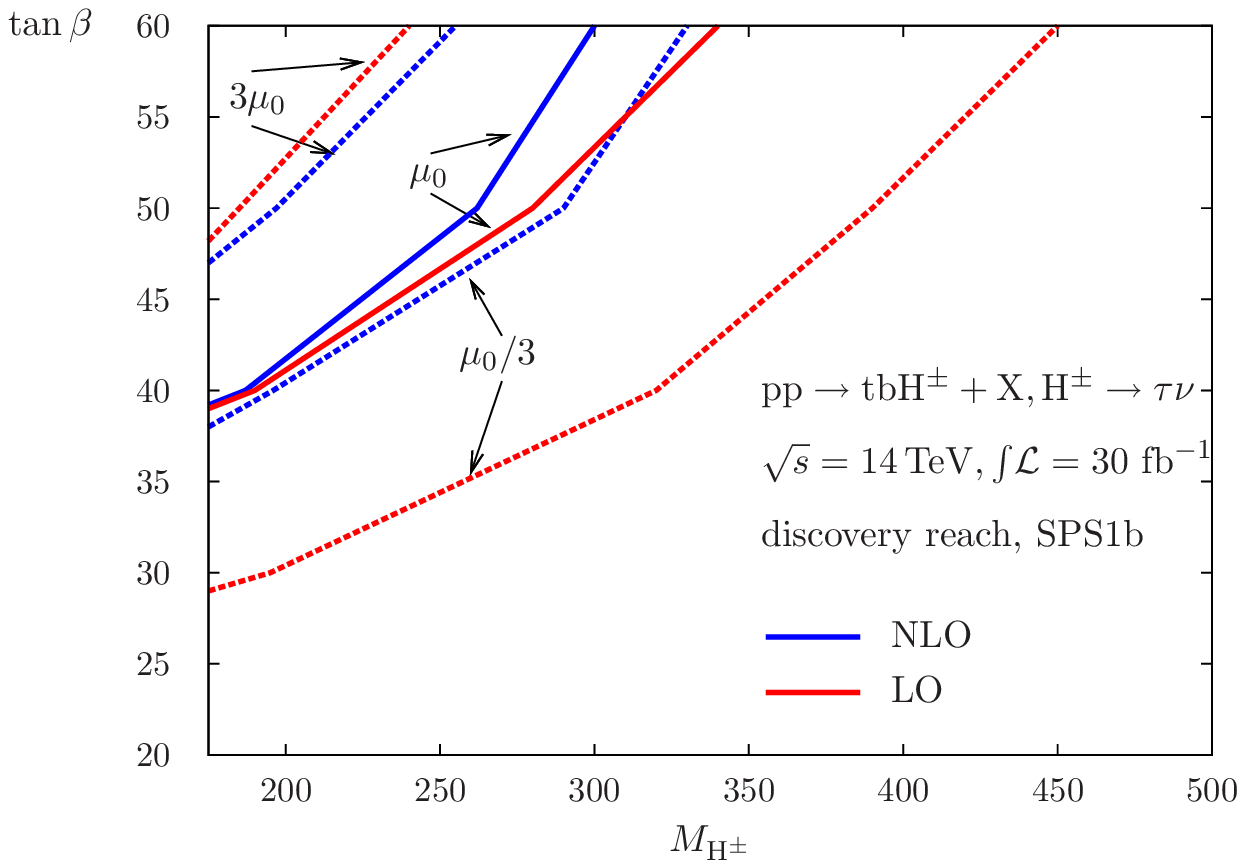,%
        bbllx=135pt,bblly=540pt,bburx=500pt,bbury=800pt,%
        scale=1.0,clip=}
\caption{Discovery reach for MSSM charged Higgs bosons ${\rm H}^\pm$,
  with ${\rm H}^\pm \to \tau\nu$, at CMS~\cite{kinnunen_2006} as a
  function of $\tan\beta$ and $M_{{\rm H}^\pm}$. All other
  supersymmetric parameters have been fixed to the SPS~1b values.
  Higgs-boson discovery with $\int\!{\cal L} = 30$~fb$^{-1}$ is
  possible in the areas above the curves.  Shown are results based on
  the LO and NLO cross sections calculation in the 4FS with the
  central scale $\mu_0 = (m_{\rm t} + m_{\Pb} + M_{{\rm H}^-})/3$ and
  scales set to $\mu = \mu_0/3$ and $3\mu_0$, respectively.}
\label{fig:discovery}
\end{figure}

Accurate theoretical predictions for the charged-Higgs production
cross section are crucial to exploit the LHC potential for MSSM
Higgs-boson searches. To exemplify the importance of reducing the
theoretical uncertainty through NLO calculations, we consider the
discovery reach in the search channel $\Pp \Pp \to {\rm tbH}^\pm+X$
followed by the hadronic decay ${\rm H}^\pm \to \tau^\pm \nu_{\tau}$
with $\tau\to {\rm hadrons} + \nu_\tau$, as analyzed for the CMS
detector in Refs.~\cite{kinnunen_2006,Hashemi:2008ma}. The number of
signal events is given by
\begin{equation}\label{eq:nev}
N_{\rm signal} = {\textstyle \int}\!{\cal L}\times\sigma(\Pp \Pp \to {\rm tbH}^\pm+X)\times
{\rm BR}({\rm H}^\pm \to \tau^\pm \nu_{\tau})\times {\rm
  BR}(\tau\to{\rm hadrons})\times {\rm exp.~efficiency}\,, 
\end{equation}
where $\int\!{\cal L}$ denotes the collider luminosity. The experimental
efficiency has been determined in Ref.~\cite{kinnunen_2006} as a
function of the Higgs-boson mass:

\begin{center}
\begin{tabular}{c|c|c|c|c|c|c}
$M_{{\rm H}^\pm}$ [GeV] & 171.6 & 180.4 & 201.0 & 300.9 & 400.7 & 600.8  \\
\hline
exp.~eff. [$10^{-4}$] & 3.5 & 4.9 & 5.0 & 23 & 32 & 42
\end{tabular}\,.
\end{center}
The QCD background processes lead to $1.7\pm 1$ events after cuts,
independent of $M_{\rm H^\pm}$, so that 14 or more signal events are
needed for a 5$\,\sigma$ discovery~\cite{kinnunen_2006}. We determine
the number of signal events from Eq.~(\ref{eq:nev}) for the benchmark
scenario SPS~1b, varying $\tan\beta$ and $M_{\rm A}$ while keeping all
other supersymmetric parameters fixed. The branching ratio ${\rm
  BR}({\rm H}^\pm \to \tau^\pm \nu_{\tau})$ varies strongly with
$M_{\rm A}$ and has been calculated with {\sl
  SUSY-Hit}~\cite{Djouadi:2006bz}. The branching ratio of the hadronic
$\tau$ decay has been set to ${\rm BR}(\tau\to{\rm hadrons}) =
0.65$~\cite{Amsler:2008zzb}, and we assume an integrated luminosity of
$\int\!{\cal L} = 30$~fb$^{-1}$.  In Figure~\ref{fig:discovery} we
show the 5$\,\sigma$ discovery contours for ${\rm H^\pm}$ as a
function of $\tan\beta$ and $M_{\rm H^\pm}$, where the number of
signal events in Eq.~(\ref{eq:nev}) has been evaluated using the LO
and NLO 4FS calculation presented in this paper.
We show results for the central scale $\mu_0 = (m_{\rm t} + m_{\Pb} +
M_{{\rm H}^-})/3$ and results for the renormalization and
factorization scales set to $\mu = \mu_0/3$ and $3\mu_0$,
respectively. Higgs-boson discovery is possible in the areas above the
curves shown in the figure.  Figure~\ref{fig:discovery} demonstrates
that the reduction of the scale uncertainty is crucial to exploit the
potential of the LHC for charged Higgs-boson discovery. Note that a
more detailed study of the supersymmetric parameter dependence of the
discovery contours is presented in Ref.~\cite{Hashemi:2008ma}.  The
importance of a reduced scale dependence through the calculation of
higher-order corrections for charged-Higgs-boson discovery, however,
is generic and largely independent of the supersymmetric scenario
considered.

\section{Conclusions}
\label{se:conclusion}

We have presented the next-to-leading order supersymmetric QCD
corrections to charged-Higgs-boson production at the LHC in the
four-flavour scheme through the parton processes $\Pq\bar \Pq,\Pg\Pg
\to {\rm t}\Pb{\rm H}^{\pm}$. While the K-factor is moderate at the
central scale $\mu = (m_{\rm t} + m_{\Pb} + M_{{\rm H}^-})/3$, the QCD
corrections considerably reduce the renormalization and factorization
scale dependence and thus stabilize the theoretical predictions. We
find that the shapes of the top-quark and Higgs transverse-momentum
distributions are not strongly affected by the higher-order
corrections. On the other hand, the bottom-quark $p_{\rm
  T}$-distribution is softened at NLO, depending in detail on the
reconstruction method of the bottom quarks.  The NLO corrections do
not significantly change the shape of the rapidity distributions at
central rapidities, but generically predict more particles at large
rapidities. We have presented a first comparison of the four-flavour
scheme NLO inclusive cross sections with a five-flavour scheme
calculation based on bottom--gluon fusion.  Even though the results
within the two schemes are consistent within the scale uncertainties,
the central predictions in the five-flavour scheme are larger than
those of the four-flavour scheme by approximately 40\%. Finally, by
referring to a recent CMS study~\cite{kinnunen_2006} we have
demonstrated that NLO predictions for the charged-Higgs production
cross section are crucial to exploit the LHC potential for MSSM
Higgs-boson searches.

\section*{Acknowledgments}
We would like to thank Tilman Plehn for providing us with the computer
code of the calculation of Ref.~\cite{Plehn:2002vy} and for
discussions on the comparison with the five-flavour scheme
calculation.  This work is
supported in part by the European Community's Marie-Curie Research
Training Network under contract MRTN-CT-2006-035505 ``Tools and
Precision Calculations for Physics Discoveries at Colliders'', the DFG
SFB/TR9 ``Computational Particle Physics'', and the Helmholtz Alliance
``Physics at the Terascale''.  We thank the Galileo Galilei Institute
for Theoretical Physics in Florence and the CERN Theory division for
their hospitality, and the INFN and CERN for partial financial
support.  The work of M.S. was further supported in part by the DOE
under Task TeV of contract DE-FGO2-96-ER40956.

\begin{appendix}

\section{SPS~1b benchmark scenario}
\label{app:SPS}

For the SPS~1b benchmark~\cite{Allanach:2002nj} scenario discussed in
this work we use the following input for $\tanb$, the supersymmetric Higgs mass parameter
$\mu$, the electroweak gaugino mass parameters $M_{1,2}$, the gluino
mass $m_{\tilde{g}}$, the trilinear couplings $A_{\tau,\Pt,\Pb}$,
the scale $\mu_R ($\DRbar$)$ at which the \DRbar\ input values are defined, 
the soft SUSY-breaking parameters in the diagonal
entries of the squark and slepton mass matrices of the first and
second generation $M_{fi}$ (where $i=L,R$ refers to the left- and
right-handed sfermions, $f=q,l$ to quarks and leptons, and $f=u,d,e$
to up and down quarks and electrons, respectively), and the
analogous soft SUSY-breaking parameters for the third generation
$M^{3G}_{fi}$:
\begin{center}
\begin{tabular}{rclrcl}
$\tanb$            & = & $  30.0$ \quad & \qquad $M_{qL}$      & = & $836.2$~GeV  \\
$\mu$              & =&$ 495.6$~GeV  \quad & \qquad $M_{dR}$           & = & $803.9$~GeV \\
$M_1$              & =&$ 162.8$~GeV \quad & \qquad $M_{uR}$           & = & $807.5$~GeV \\
$M_2$              & =&$ 310.9$~GeV \quad & \qquad $M_{lL}$           & = & $334.0$~GeV \\
$m_{\tilde{g}}$    & =&$ 916.1$~GeV \quad & \qquad $M_{eR}$           & = & $248.3$~GeV \\
$A_\tau$           & =&$-195.8$~GeV \quad & \qquad $M^{3G}_{qL}$      & = & $762.5$~GeV \\
$A_\Pt$            & =&$-729.3$~GeV \quad & \qquad $M^{3G}_{dR}$      & = & $780.3$~GeV \\
$A_\Pb$            & =&$-987.4$~GeV \quad & \qquad $M^{3G}_{uR}$      & = & $670.7$~GeV \\
$\mu_R ($\DRbar$)$ & =&$ 706.9$~GeV \quad & \qquad $M^{3G}_{lL}$      & = & $323.8$~GeV \\
$M^{3G}_{eR}$      & =& $218.6$~GeV\,. & & 
\end{tabular}
\end{center}
The mass of the CP-odd Higgs boson $\MA$ is varied and taken as input
to calculate the charged-Higgs boson mass $M_{{\rm H}^\pm}$, taking
into account higher-order corrections up to two loops in the effective
potential approach~\cite{Carena:1995bx, Haber:1996fp} as included in
the program {\sl HDECAY}~\cite{Djouadi:1997yw}.

\end{appendix}


\clearpage

 
\begin{thebibliography}{99}
\frenchspacing

\bibitem{Higgs:1964ia}
P.~W.~Higgs,
Phys.\ Lett.\  {\bf 12}, 132 (1964);
Phys.\ Rev.\ Lett.\  {\bf 13}, 508 (1964)
and Phys.\ Rev.\  {\bf 145}, 1156 (1966);
F.~Englert and R.~Brout,
Phys.\ Rev.\ Lett.\  {\bf 13}, 321 (1964);
G.~S.~Guralnik, C.~R.~Hagen and T.~W.~Kibble,
Phys.\ Rev.\ Lett.\  {\bf 13}, 585 (1964).

\bibitem{Heister:2002ev}
  A.~Heister {\it et al.}  [ALEPH Collaboration],
  Phys.\ Lett.\  B {\bf 543} (2002) 1
  [arXiv:hep-ex/0207054].

\bibitem{Gunion:1988pc}
  J.~F.~Gunion and A.~Turski,
  Phys.\ Rev.\  D {\bf 39} (1989) 2701.

\bibitem{Brignole:1991wp}
  A.~Brignole,
  Phys.\ Lett.\  B {\bf 277} (1992) 313.

\bibitem{Diaz:1991ki}
  M.~A.~Diaz and H.~E.~Haber,
  Phys.\ Rev.\  D {\bf 45} (1992) 4246.

\bibitem{Frank:2006yh}
  M.~Frank, T.~Hahn, S.~Heinemeyer, W.~Hollik, H.~Rzehak and G.~Weiglein,
  JHEP {\bf 0702} (2007) 047
  [arXiv:hep-ph/0611326].

\bibitem{Schael:2006cr}
  S.~Schael {\it et al.}  [ALEPH Collaboration],
  Eur.\ Phys.\ J.\  C {\bf 47} (2006) 547
  [arXiv:hep-ex/0602042].

\bibitem{Abulencia:2005jd}
  A.~Abulencia {\it et al.}  [CDF Collaboration],
  Phys.\ Rev.\ Lett.\  {\bf 96} (2006) 042003
  [arXiv:hep-ex/0510065].

\bibitem{Abazov:2001md}
  V.~M.~Abazov {\it et al.}  [D0 Collaboration],
  Phys.\ Rev.\ Lett.\  {\bf 88} (2002) 151803
  [arXiv:hep-ex/0102039].

\bibitem{:1999fq} 
  ATLAS: Detector and physics performance technical
  design report, CERN-LHCC-99-014.

\bibitem{Ball:2007zza}
  G.~L.~Bayatian {\it et al.}  [CMS Collaboration],
  J.\ Phys.\ G {\bf 34} (2007) 995.

\bibitem{Djouadi:2005gj}
  A.~Djouadi,
  Phys.\ Rept.\  {\bf 459} (2008) 1
  [arXiv:hep-ph/0503173].

\bibitem{Dittmaier:2007uw}
  S.~Dittmaier, G.~Hiller, T.~Plehn and M.~Spannowsky,
  Phys.\ Rev.\  D {\bf 77} (2008) 115001
  [arXiv:0708.0940 [hep-ph]].

\bibitem{Barnett:1987jw}
  R.~M.~Barnett, H.~E.~Haber and D.~E.~Soper,
  Nucl.\ Phys.\ B {\bf 306} (1988) 697;\\
  D.~A.~Dicus and S.~Willenbrock,
  Phys.\ Rev.\ D {\bf 39} (1989) 751.

\bibitem{Dittmaier:2003ej}
  S.~Dittmaier, M.~Kr\"amer and M.~Spira,
  Phys.\ Rev.\ D {\bf 70} (2004) 074010
  [hep-ph/0309204].

\bibitem{Campbell:2004pu}
  J.~Campbell {\it et al.},
  [hep-ph/0405302].

\bibitem{Dawson:2005vi}
  S.~Dawson, C.~B.~Jackson, L.~Reina and D.~Wackeroth,
  Mod.\ Phys.\ Lett.\ A {\bf 21} (2006) 89
  [hep-ph/0508293].

\bibitem{Buttar:2006zd}
  C.~Buttar {\it et al.},
  [hep-ph/0604120].

\bibitem{Zhu:2001nt}
  S.~h.~Zhu,
  Phys.\ Rev.\  D {\bf 67} (2003) 075006
  [arXiv:hep-ph/0112109].

\bibitem{Gao:2002is}
  G.~P.~Gao, G.~R.~Lu, Z.~H.~Xiong and J.~M.~Yang,
  Phys.\ Rev.\  D {\bf 66} (2002) 015007
  [arXiv:hep-ph/0202016].

\bibitem{Plehn:2002vy}
  T.~Plehn,
  Phys.\ Rev.\  D {\bf 67} (2003) 014018
  [arXiv:hep-ph/0206121].

\bibitem{Berger:2003sm}
  E.~L.~Berger, T.~Han, J.~Jiang and T.~Plehn,
  Phys.\ Rev.\  D {\bf 71} (2005) 115012
  [arXiv:hep-ph/0312286].

\bibitem{Kidonakis:2005hc}
  N.~Kidonakis,
  PoS {HEP2005} (2006) 336 [arXiv:hep-ph/0511235].

\bibitem{Peng:2006wv}
  W.~Peng, M.~Wen-Gan, Z.~Ren-You, J.~Yi, H.~Liang and G.~Lei,
  Phys.\ Rev.\  D {\bf 73} (2006) 015012
  [arXiv:hep-ph/0601069].

\bibitem{kinnunen_2006}
  R.~Kinnunen,
  CMS NOTE-2006/100.

\bibitem{DiazCruz:1992gg}
  J.~L.~Diaz-Cruz and O.~A.~Sampayo,
  Phys.\ Rev.\  D {\bf 50} (1994) 6820.

\bibitem{Borzumati:1999th}
  F.~Borzumati, J.~L.~Kneur and N.~Polonsky,
  Phys.\ Rev.\  D {\bf 60} (1999) 115011
  [arXiv:hep-ph/9905443].

\bibitem{Miller:1999bm}
  D.~J.~Miller, S.~Moretti, D.~P.~Roy and W.~J.~Stirling,
  Phys.\ Rev.\  D {\bf 61} (2000) 055011
  [arXiv:hep-ph/9906230].

\bibitem{Yao:2006px}
  W.~M.~Yao {\it et al.}  [Particle Data Group],
  J.\ Phys.\ G {\bf 33} (2006) 1.

\bibitem{Beenakker:2001rj}
  W.~Beenakker, S.~Dittmaier, M.~Kr\"amer, B.~Pl\"umper, M.~Spira and P.~M.~Zerwas,
  Phys.\ Rev.\ Lett.\  {\bf 87} (2001) 201805
  [arXiv:hep-ph/0107081].

\bibitem{Beenakker:2002nc}
  W.~Beenakker, S.~Dittmaier, M.~Kr\"amer, B.~Pl\"umper, M.~Spira and P.~M.~Zerwas,
  Nucl.\ Phys.\  B {\bf 653} (2003) 151
  [arXiv:hep-ph/0211352].

\bibitem{Dawson:2002tg}
  S.~Dawson, L.~H.~Orr, L.~Reina and D.~Wackeroth,
  Phys.\ Rev.\  D {\bf 67} (2003) 071503
  [arXiv:hep-ph/0211438].

\bibitem{diss_walser} 

  Manuel Walser, ``NLO QCD and SUSY-QCD
  corrections to associated MSSM Higgs production with heavy quarks at
  hadron colliders'', disseration ETH Z\"urich NO. 17592, 2008

\bibitem{Kublbeck:1990xc}
J.~K\"ublbeck, M.~B\"ohm and A.~Denner,
Comput.\ Phys.\ Commun.\  {\bf 60} (1990) 165;
\\
H.~Eck and J.~K\"ublbeck, {\it Guide to FeynArts 1.0\/},
University of W\"urzburg, 1992.

\bibitem{Hahn:2000kx}
T.~Hahn,
Comput.\ Phys.\ Commun.\  {\bf 140} (2001) 418
[hep-ph/0012260].


\bibitem{Dittmaier:2003bc}
  S.~Dittmaier,
  Nucl.\ Phys.\  B {\bf 675} (2003) 447
  [arXiv:hep-ph/0308246].

\bibitem{Bredenstein:2008zb}
  A.~Bredenstein, A.~Denner, S.~Dittmaier and S.~Pozzorini,
  JHEP {\bf 0808} (2008) 108
  [arXiv:0807.1248 [hep-ph]].

\bibitem{Denner:2002ii}
  A.~Denner and S.~Dittmaier,
  Nucl.\ Phys.\ B {\bf 658} (2003) 175
  [hep-ph/0212259].

\bibitem{Passarino:1978jh}
G.~Passarino and M.~Veltman,
Nucl.\ Phys.\ B {\bf 160} (1979) 151.

\bibitem{Denner:2005nn}
  A.~Denner and S.~Dittmaier,
  Nucl.\ Phys.\  B {\bf 734} (2006) 62
  [arXiv:hep-ph/0509141].

\bibitem{'tHooft:1978xw}
G.~'t Hooft and M.~Veltman,
Nucl.\ Phys.\ B {\bf 153} (1979) 365;\\
%
W.~Beenakker and A.~Denner,
Nucl.\ Phys.\ B {\bf 338} (1990) 349;\\
%
A.~Denner, U.~Nierste and R.~Scharf,
Nucl.\ Phys.\ B {\bf 367} (1991) 637.


\bibitem{Hahn:1998yk}
  T.~Hahn and M.~Perez-Victoria,
  Comput.\ Phys.\ Commun.\  {\bf 118} (1999) 153
  [arXiv:hep-ph/9807565];
  G.~J.~van Oldenborgh,
  Comput.\ Phys.\ Commun.\  {\bf 66} (1991) 1.

\bibitem{Catani:1996vz}
  S.~Catani and M.~H.~Seymour,
  Nucl.\ Phys.\  B {\bf 485} (1997) 291
  [Erratum-ibid.\  B {\bf 510} (1998) 503]
  [arXiv:hep-ph/9605323];\\
%
  S.~Catani, S.~Dittmaier, M.~H.~Seymour and Z.~Tr\'ocs\'anyi,
  Nucl.\ Phys.\  B {\bf 627} (2002) 189
  [arXiv:hep-ph/0201036].

\bibitem{Alwall:2007st}
  J.~Alwall {\it et al.},
  JHEP {\bf 0709} (2007) 028
  [arXiv:0706.2334 [hep-ph]].

\bibitem{Murayama:1992gi}
  H.~Murayama, I.~Watanabe and K.~Hagiwara,
  KEK-91-11, January 1992 (unpublished).

\bibitem{Braaten:1980yq}
  E.~Braaten and J.~P.~Leveille,
  Phys.\ Rev.\ D {\bf 22} (1980) 715;\\
%
  M.~Drees and K.~I.~Hikasa,
  Phys.\ Lett.\ B {\bf 240} (1990) 455
  [Erratum-ibid.\ B {\bf 262} (1991) 497].

\bibitem{Hall:1993gn}
  L.~J.~Hall, R.~Rattazzi and U.~Sarid,
  Phys.\ Rev.\ D {\bf 50} (1994) 7048
  [hep-ph/9306309].

\bibitem{Hempfling:1993kv}
  R.~Hempfling,
  Phys.\ Rev.\ D {\bf 49} (1994) 6168.

\bibitem{Carena:1994bv}
  M.~Carena, M.~Olechowski, S.~Pokorski and C.~E.~M.~Wagner,
  Nucl.\ Phys.\ B {\bf 426} (1994) 269
  [hep-ph/9402253].

\bibitem{Pierce:1996zz}
  D.~M.~Pierce, J.~A.~Bagger, K.~T.~Matchev and R.~J.~Zhang,
  Nucl.\ Phys.\ B {\bf 491} (1997) 3
  [hep-ph/9606211].

\bibitem{Carena:1999py}
  M.~Carena, D.~Garcia, U.~Nierste and C.~E.~M.~Wagner,
  Nucl.\ Phys.\ B {\bf 577} (2000) 88
  [hep-ph/9912516].

\bibitem{Guasch:2003cv}
  J.~Guasch, P.~H\"afliger and M.~Spira,
  Phys.\ Rev.\ D {\bf 68} (2003) 115001
  [hep-ph/0305101].

\bibitem{Group:2008nq}
    [CDF Collaboration],
  arXiv:0803.1683 [hep-ex].

\bibitem{Allanach:2002nj}
  B.~C.~Allanach {\it et al.},
  Eur.\ Phys.\ J.\ C {\bf 25} (2002) 113
  [eConf {\bf C010630} (2001) P125]
  [hep-ph/0202233].
 
\bibitem{Carena:1995bx}
M.~Carena, H.E.~Haber, S.~Heinemeyer, W.~Hollik, C.E. Wagner and
G.~Weiglein,
Nucl.~Phys.~{\bf B580} (2000) 29.

\bibitem{Haber:1996fp}
  H.~E.~Haber, R.~Hempfling and A.~H.~Hoang,
  Z.\ Phys.\  C {\bf 75} (1997) 539
  [arXiv:hep-ph/9609331].

\bibitem{Djouadi:1997yw}
  A.~Djouadi, J.~Kalinowski and M.~Spira,
  Comput.\ Phys.\ Commun.\  {\bf 108} (1998) 56
  [arXiv:hep-ph/9704448].

\bibitem{Martin:2004ir}
  A.~D.~Martin, R.~G.~Roberts, W.~J.~Stirling and R.~S.~Thorne,
  Phys.\ Lett.\  B {\bf 604} (2004) 61
  [arXiv:hep-ph/0410230].

\bibitem{Noth:2008tw}
  D.~Noth and M.~Spira,
  Phys.\ Rev.\ Lett.\  {\bf 101} (2008) 181801
  [arXiv:0808.0087 [hep-ph]].

\bibitem{Martin:2006qz}
  A.~D.~Martin, W.~J.~Stirling and R.~S.~Thorne,
  Phys.\ Lett.\  B {\bf 636} (2006) 259
  [arXiv:hep-ph/0603143].

\bibitem{Gluck:2008gs}
  M.~Gl\"uck, P.~Jimenez-Delgado, E.~Reya and C.~Schuck,
  Phys.\ Lett.\  B {\bf 664} (2008) 133
  [arXiv:0801.3618 [hep-ph]].

\bibitem{Dittmaier:2006cz}
  S.~Dittmaier, M.~Kr\"amer, A.~M\"uck and T.~Schl\"uter,
  JHEP {\bf 0703} (2007) 114
  [arXiv:hep-ph/0611353].

\bibitem{Hashemi:2008ma}
  M.~Hashemi, S.~Heinemeyer, R.~Kinnunen, A.~Nikitenko and G.~Weiglein,
  arXiv:0804.1228 [hep-ph].

\bibitem{Djouadi:2006bz}
  A.~Djouadi, M.~M.~M\"uhlleitner and M.~Spira,
  Acta Phys.\ Polon.\  B {\bf 38} (2007) 635
  [arXiv:hep-ph/0609292].

\bibitem{Amsler:2008zzb}
  C.~Amsler {\it et al.}  [Particle Data Group],
  Phys.\ Lett.\  B {\bf 667} (2008) 1.


\end{thebibliography}
\end{document}